\documentclass[12pt,preprint]{aastex}

\newcommand{\ha}{H$\alpha$}
\newcommand{\hb}{H$\beta$}

\newcommand{\teff}{$T_{\mathrm{eff}}$}

\newcommand{\cho}{{\sc\footnotesize  CHORIZOS}}
\newcommand{\stb}{{\sc\footnotesize Starburst99}}
\newcommand{\rv}{$R_{5495}$}
\newcommand{\ecc}{$E(4405-5495)$}
  
\newcommand{\hrd}{ Hertzsprung--Russell diagram}
\newcommand{\mbol}{$M_\mathrm{bol}$}
\slugcomment{Published by The Astronomical Journal}
\shorttitle{The young stellar population of NGC 4214.II}
\shortauthors{\'Ubeda et al.}

\begin{document}

\title{The young stellar population of NGC 4214 as observed with 
 HST. II. Results.\footnote{Based on observations made with the NASA/ESA 
{\it Hubble Space Telescope,} obtained at the Space Telescope Science
 Institute, which is operated by the Association of Universities for 
 Research in Astronomy, Inc., under NASA contract NAS 5-26555.}}

\author{ Leonardo \'{U}beda, Jes\'us  
Ma\'{i}z Apell\'aniz\footnote{Affiliated with the Space Telescope Division of the 
European Space Agency, ESTEC, Noordwijk, Netherlands.} , \& John W. MacKenty}
\email{lubeda@stsci.edu, jmaiz@stsci.edu, mackenty@stsci.edu}
\affil{Space Telescope Science Institute, 3700 San 
Martin Drive, Baltimore, MD 21218, U.S.A.}

\begin{abstract} 
We present the results of a detailed UV--optical study of the
nearby dwarf starburst galaxy NGC~4214 using multifilter HST/WFPC2+STIS photometry.
The stellar extinction is found to be quite patchy, with some areas
having values of $\ecc < 0.1$ mag and others, associated with star forming regions, much more heavily obscured, 
a result which is consistent with previous studies of the nebular extinction. We determined the ratio of
blue--to--red supergiants and found it to be consistent with theoretical models for the metallicity of the SMC.
The stellar IMF of the field in the range $20-100 \,\,M_\odot$ is found to be steeper than $\gamma=-2.8$
 ($\gamma=-2.35$  for a Salpeter IMF). 
A number of massive clusters and associations  with ages between a few
and 200 million years are detected and their properties are discussed.
\end{abstract}

\keywords{galaxies: individual (\objectname{NGC~4214}) --- galaxies: star clusters ---
galaxies: stellar content --- stars: early-type --- 
stars: luminosity function, mass function}

\section{Introduction}
In  this paper we present the results of
a comprehensive study of NGC~4214. In \cite{Ubedetala06}  (hereafter Paper~I) we 
presented and discussed the deep UV--optical images from two HST instruments:
WFPC2 and STIS.
We described  the methods that we had employed to 
reduce and analyze our HST images, and we briefly discussed  its stellar populations
by means of the analysis of 
two
Hertzsprung--Russell   [ $\log(T_{\mathrm{eff}}) $, $M_{\mathrm{bol}}$ ]  diagrams. 

In Paper~I we made a thorough   description of 
how we used \cho\  ~\citep{Maiz04},   
an IDL package
that fits an arbitrary family of spectral energy  distribution 
models (SEDs),  to analyze our multi--color photometric data. 
We applied \cho\ to stars and clusters separately, using different 
input SEDs. In this present work we show and discuss  the results.

NGC~4214 (see Figures 1, 2 and 4 in Paper~I)  is a nearby ($2.94 \pm 0.18 $ Mpc; \cite{Maizetal02a})  dwarf 
galaxy \citep{Vauc91} with a low metallicity content $(Z=0.006)$  \citep{Ubedetala06}.
It has two main regions of star formation which we studied using 
high--resolution images obtained from three HST archival proposals:
6716 (P.I.: Theodore Stecher), 6569  (P.I.: John MacKenty), and
9096 (P.I.: Jes\'us Ma\'{i}z-Apell\'aniz).

NGC~4214 possesses a combination of properties (high star formation rate, proximity, 
low extinction, spatially well resolved) that  make it a perfect candidate  for astrophysical studies
of  young stellar populations,
which we investigate in the present work. 
We analyze the ratio of blue to red (B/R)
supergiants, the initial stellar mass function (IMF), and the properties of its young-- and
intermediate--age cluster population. We also study the variable extinction across the galaxy.

In order to advance the understanding of massive stars
it is important to have an observational database with which the 
predictions of stellar evolutionary theory may be compared and refined.
In particular, the ratio B/R of the blue to red supergiants is a major characteristic of the luminous 
star population in galaxies.
\Citet{Bergh68}  first suggested the B/R ratio varied among nearby galaxies as a 
result of the effect of metallicity on massive star evolution. Several studies confirmed that 
the B/R ratio is  an increasing function of metallicity   \citep{LaMa94,Eggeetal02}. This
fact is known as  a result of studies in the Galaxy \citep{HumpMcel84} and  the Magellanic Clouds
\citep{Hump83,MassOlse03}.  
This ratio has also been studied in M33  \citep{Walk64,HumpSand80} and more recently 
in Sextans A by \cite{DohmSkil02}.  In this paper we examine the B/R ratio    for NGC~4214.

\cite{Krou02} presents an updated review
of our current knowledge of the IMF for different types of stars.
This function is particularly important for the most
massive stars \citep{Garmetal82}    
which evolve very quickly and strongly influence their environment via stellar winds and 
mass loss.  \cite{Massetal95, Massetal95b} studied the massive star content of 
OB associations in the Milky Way and the Magellanic Clouds 
and found that there is no difference in the IMF slope between the MW, the LMC
and the SMC; therefore, to first order, metallicity does not appear to affect the IMF of massive stars, 
at least over the factor of 5 spanned
between the SMC ($Z=0.004$), the LMC ($Z=0.008$)  and the MW ($Z=0.02$).
OB associations have  been discovered in some Local Group galaxies, and with the 
fine spatial resolution provided by the HST, it is possible to obtain the stellar IMF
of those systems. Recent works include M31  \citep{Veltetal04}, M33  \citep{GonDPere00}, 
IC10 \citep{Hunt01}, and  IC1613  \citep{Georetal99}. The previous work on the IMF of 
NGC~4214 has been based on the integrated spectrum of the galaxy 
\citep{Leitetal96,Chanetal05}. Here we present the first work on its IMF based on high--resolution
HST imaging.

Stellar clusters have long been recognized as important laboratories for astrophysical
research. They are extremely useful in different aspects of Astronomy: they
provide classical tests of stellar evolution, they can be used for studies of stellar dynamics
and they provide an
understanding of the galactic structure \citep{LadaLada03}.
Young clusters are tracers of recent star formation; they form in 
giant molecular clouds (GMCs) and they remain embedded or close to
their parent cloud during 
the first $\approx 1$ Myr, making it difficult to study them in their early stages.
However, recent developments in infrared imaging cameras have led 
to the conclusion that young embedded clusters are numerous, and that
a significant fraction of all stars may form in such systems. 
Massive Young Clusters (MYCs) can be divided into 
Super Star Clusters (SSCs), which are organized  
around a compact core (size $\approx 1-3$ pc), and 
Scaled OB Associations (SOBAs) which lack such a
structure and are more extended objects (size $> 10$ pc). 
SSCs are bound objects and represent the high--mass 
end of young stellar clusters while SOBAs are 
unbound and are the massive relatives of regular OB associations \citep{Maiz01b}.  
MYCs have been discovered in various environments, including our Galactic Center,
in the nuclei of late--type galaxies, in nearby starburst galaxies and  in merging galaxies.

The present paper is organized as follows: In Section 2  we present 
the study of the extinction throughout  NGC~4214; in Section 3 we analyze the B/R ratio;
the IMF is addressed in Section 4; in Section 5 we perform a detailed  study of 
some interesting clusters. 
Finally, in Section 6 we summarize our results and provide the conclusions.

\section{Extinction}

 A comparison between stellar atmosphere models and 
 observed colors can be used to infer the extinction toward 
individual stars. For our study, we used the IDL code 
  \cho ~\citep{Maiz04};
this program (see Section 2.5 in Paper~I) 
reads unreddened SED models, extinguishes them and obtains the 
synthetic photometry
that is compatible with  the input colors.    In this way
we obtained the extinction law  that best agrees  with the data,
and the extinction value   [ \ecc ~] ~for each star in our main photometric
lists:  {\sc\footnotesize  LIST336} (built with F336W as reference filter) and
{\sc\footnotesize  LIST814}  (built with F814W as reference filter) as described in Paper~I.
For the extinction law, we considered  
the $R_{5495}-$dependent family of 
 \citet{Cardetal89} (Galactic extinction); the average LMC and 
LMC2 laws of  \citet{Missetal99}   (extinction of the LMC), and the SMC
law of  \citet{GordClay98}  (extinction of the SMC).  \cho\ also provides
the uncertainties $\sigma  $  of the calculated parameters. 

In  previous studies of NGC~4214, the extinction correction 
was performed using different values of  $E(B-V)$:  \citet{Maizetal98} 
measured a variable $E(B-V)$ with values between 0.0 and 0.6 mag from the ratio of the nebular emission
lines \ha\ to \hb. \citet{Maizetal02a} measured an average value of $E(B-V)=0.09$ mag from the optical
colors of the young stellar population far from the main star--forming complexes. \citet{Droz02} and 
\citet{Calzetal04} adopted the value $E(B-V)=0.02$ mag provided by the IRAS DIRBE map 
of \citet{Schletal98}. All those estimates assume a \citet{Cardetal89} extinction law with $R_{5495}=3.1$.

\begin{figure}
\includegraphics[width=0.9\linewidth]{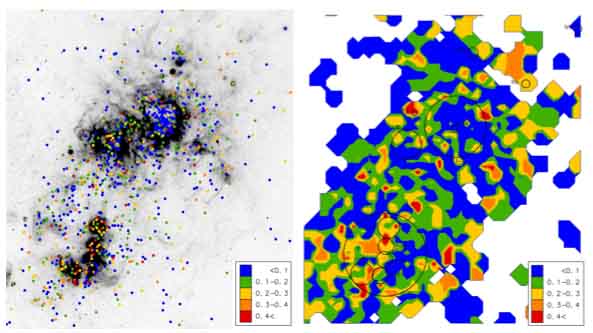}
\caption{ 
 [Left panel] F656N mosaic of NGC~4214 where we have over--plotted the selection of
 stars that we used to make the extinction map. The color scale represents 
 values of \ecc\
 [Right panel] Extinction map of the central region of NGC~4214. 
 The  orientation  is north pointing up and east pointing to the left. 
The field dimensions are  875 pc $\times$ 972 pc or  $ 61 \farcs 4 \times 68 \farcs 3$.
 The cluster apertures are also
 drawn for reference. See the text for a description of
 how  we made this map.
}
 \label{fig01}
\end{figure}

In order to develop an extinction map of NGC~4214, we made a selection of
objects from  {\sc\footnotesize  LIST336}, using the following criteria:
We considered all the objects with
$\sigma_{E(4405-5495)}  \leqslant 0.1 $ mag  for those stars with $E(4405-5495)  \leqslant 0.4 $ mag.
This selection yields 855 objects with $<{\mathrm F336W}>=20.93$ and  $<\sigma_{\mathrm  F336W}>=1.22$.
Objects with  $E(4405-5495)  > 0.4 $  mag
 were selected on the basis of their F336W magnitude 
 (F336W $< 21$ mag)  and their location within the galaxy.
 We made no restriction according to photometric or
 extinction error in this case. We found 43 objects with $E(4405-5495)  > 0.4 $
 and they present $<E(4405-5495) >= 0.54$ mag and 
 $<\sigma_{E(4405-5495)} > = 0.21 $ mag.
%
Using this biased sample of objects, we
 may be underestimating the extinction values and probably 
missing some blue objects.
It is important to note that extinction is a complicated three-dimensional effect which 
involves stars, nebulae and dust and which varies at  small  ($  \lesssim 1$ pc)  scales.
All these facts contribute  to  make extinction--modelling 
quite difficult.
Finally, we added the results of the extinction analysis (Section 5.1)
for clusters I--As, I--Es, IIIs, and IVs to our list. 
The left panel of Figure~\ref{fig01} is an F656N  mosaic of the surveyed region of NGC~4214,
where we have marked the  objects in our final list with a color scheme
to indicate the extinction   distribution throughout the galaxy.  
To build the \ha\ (filter F656N) mosaic, we used  images u3n8010fm + gm from
proposal 6569. The total exposure time is 1600 sec. This mosaic clearly shows the patchy 
nature of the extinction. 
 
To build the extinction map, we created a  spatial grid $M$ and 
assigned a weighted value of the extinction
to each pixel in the grid. The procedure follows:
We first calculated the distance $d_{ijk}$ (in  WF pixels of $0\farcs1$) between each pixel in the grid $M[i,j]$
and all the stars $k$ in our list.  We then 
calculated a weight with
\begin{equation} 
w_k = \frac{\exp\left( - \frac{ d_{ijk}^2}{30}    \right)}{\sigma_{E_k}^2}  ,
\end{equation}

\noindent where  $\sigma_{E_k}$  is the error in $ E =E(4405-5495) $ for star $k$, and
the value 30 is a scale factor. 
We used several values of the scale factor to represent the extinction map, and 
we present here the map with the one  that best showed the different extinction  regions. 
The value associated 
to the pixel $M[i,j]$ is given by 

\begin{equation}
\label{eqn-0}  
M[i,j] = \frac{\sum_k E_k \cdot w_k}{\sum_k w_k} 
\end{equation}

The extinction map is shown in the right panel
of Figure~\ref{fig01}. The most prominent feature in this map is the fact that
NGC~4214 is characterized by low values of the extinction, except for some well defined
regions with high values,  which is in agreement
with   \cite{Maizetal98}, \cite{Droz02}, and  \cite{Calzetal04}.
In Section 5  we discuss in more detail the regions around some of the clusters.

\cite{Maizetal98} and \citet{Maiz00} used the Balmer ratio (H$\alpha$/H$\beta$) as a tracer of the
reddening that affects the ionized gas and produced maps of this ratio.
Their analysis showed a significant difference between the two most prominent 
complexes in the galaxy: NGC~4214--I and  NGC~4214--II. Their conclusions are
that the nebular emission and stellar continuum are produced in co--spatial 
or close regions in NGC~4214--II, while the emitting gas is clearly spatially offset with 
respect to the stellar cluster in  the brightest knots on the NGC~4214--I complex. They also find 
that the reddening in NGC 4214--II is, on average, higher than in NGC 4214--I. Our results, derived
from the stellar colors, validate those points. The two main cavities in NGC 4214--I show low 
extinction surrounded by higher values while for NGC 4214--II the extinction is higher overall.

\cite{Waltetal01} present an interferometric study of the molecular gas in NGC~4214. They
detect three regions of molecular emission, in the northwest, southeast, and center of 
the galaxy. 
These authors compared the structure of  the molecular tracer, CO, with
tracers of star formation like \ha.
Two of the three CO complexes are associated directly with star--forming regions.
The southeastern CO clump appears co--spatial  with  NGC~4214--II. The peak 
of the CO emission is almost on top of one of the clusters. Our extinction map shows 
a high extinction region co--spatial with this molecular cloud. 
The central CO emission is associated with the largest region in \ha\ emission, 
spanning most of the NGC~4214--I region. 
This CO complex is diffuse instead of centrally concentrated, and the peak 
of the CO emission is shifted to the west of the peak of the \ha\ emission, with little CO  seen at
the location of the two main cavities. Overall, our extinction map traces the molecular cloud in this 
part of the galaxy too, although with less detail. 
 
 The extinction derived from the stellar continuum  
is similar to the extinction derived via the analysis of 
nebular lines across the galaxy, and this is true on a star by star basis.
The coincidence is fairly good throughout the galaxy.

\section{The ratio of blue--to--red supergiants }

The ratio B/R of the blue to red supergiants of initial masses
 larger than 
15 $M_{\sun}$    is an important observable  of the luminous 
star population in galaxies and it is one of the stellar properties that can
be easily measured beyond the Local Group. 
This quantity depends  strongly on the model parameters 
and it can be used to constrain the model physics very accurately. This ratio has
been calculated in Galactic clusters and  in some clusters in the LMC and SMC. 
Both \cite{Eggeetal02} and  \cite{LaMa94}  present comprehensive reviews of past studies.  

The main result from previous observational studies 
 is that, for a given luminosity range, B/R steeply increases with 
increasing metallicity $Z$, by a factor of about 10 between the SMC and the inner 
Galactic regions. Current theoretical models of massive stars are unable to 
correctly reproduce the changes of B/R with metallicity, from solar to SMC 
value. It is known that the B/R ratio is a sensitive quantity  to mass loss,
rotation, convection and mixing processes, hence it constitutes  an important 
and sensitive test for stellar evolution models if it were fully understood.

In this paper we  present our results for NGC~4214. 
The B/R value is dependent on how stars are counted, and thus disagreement
with the predictions of stellar evolutionary models have to be carefully evaluated. 
Notice that the definition of B/R is not always the same. We followed the method 
suggested  by  \cite{MaedMeyn01}. We count in the B/R ratio the B star 
models from the end of the main sequence to type B9.5 I, which corresponds 
to $\log(T_{\mathrm{eff}})  =3.99 $ according to the calibration by \cite{Flow96}.
We count as red supergiants all star models below  $\log(T_{\mathrm{eff}})  = 3.70 $. 
In both cases we considered stars located between the 15 and 25   $M_{\sun}$
evolutionary  tracks.
Figures~8 and 9 in Paper~I  show the detailed evolutionary tracks 
 of non--rotating stellar models for initial 
masses between 5, 7, 10, 12, 15, 20, 25, 40, and 120   $M_{\sun}$,  LMC--like metallicity $(Z=0.008) $
from \cite{Schaetal93}.
Two vertical lines (at $\log(T_{\mathrm{eff}})  =3.99 $ and
$\log(T_{\mathrm{eff}})  =3.70 $) mark the limits between  the blue 
supergiants locus and the position of the red supergiants.
The  location of the blue and  red supergiants is marked in those
Figures with two  polygons  
between the 15 and 25   $M_{\sun}$ tracks.
We used Figure 8 in Paper~I to count red supergiants and Figure~9 
in Paper~I  to count blue supergiants.

Several studies identify  supergiants by considering objects brighter than  
$M_{\mathrm{bol}} = -7.5 $ mag  which corresponds to masses larger than 
15 $M_{\sun}$  for red supergiants. If a lower luminosity is chosen there is a 
chance of contamination by intermediate asymptotic giant branch (AGB) stars 
  \citep{Brunetal86}.
We avoided this problem by selecting objects located above
the 15 $M_{\sun}$ evolutionary track in all cases.
We  performed an analysis to study how the  
stars located below the 15 $M_{\sun}$ evolutionary track
would contaminate  those which we consider red supergiants.
For this contamination analysis we generated artificial stars with masses that
place them below the 15 $M_{\sun}$ evolutionary track and used the typical uncertainties 
obtained from \cho\ for  \teff\ and \mbol\ to modify their positions in the H--R diagram accordingly. 
Then, using the observed distribution below the 15 $M_{\sun}$ track, we tested whether a significant 
number could have ``leaked'' to a higher mass bin due to the experimental uncertainties. 
The contamination thus calculated was found to be less than 1\%, so a correction was not applied.
For our analysis, we assumed that we are considering single stars and therefore no correction for 
blends of single stars was performed. 

In order to compare our observational results with what theory predicts, 
we calculated the   B/R   ratio values using three grids of theoretical models
from the Geneva database: [1] the evolutionary tracks of non--rotating 
($v_{ini} = 0$ km~s$^{-1}$) stellar models for initial masses between 
9 and 60   $M_{\sun}$ and SMC--like metallicity $(Z=0.004) $  \citep{MaedMeyn01}; 
[2]    the evolutionary tracks of rotating ($v_{ini} = 300$ km~s$^{-1}$)
stellar models for initial masses between 9 and 60   $M_{\sun}$ 
and SMC--like metallicity $(Z=0.004)  $   \citep{MaedMeyn01}; and  [3]   
the evolutionary tracks of non--rotating stellar models for initial 
masses between 10 and 120   $M_{\sun}$,  LMC--like metallicity $(Z=0.008) $
\citep{Schaetal93}. We did not find any major difference between models
with high and normal mass--loss rates, so we adopted normal mass--loss
rates evolutionary tracks. 
Unfortunately, evolutionary tracks of rotating ($v_{ini} = 300$ km~s$^{-1}$)
stellar models with LMC--like metallicity are not available in the literature.

The ratio of the densities of blue  supergiants to red supergiants at a given initial
mass is equal to the ratio of the lifetimes along the evolutionary tracks in the
corresponding $\log(T_{\mathrm{eff}})  $  intervals. The 
resulting  theoretical B/R values are given in  Table~\ref{tbl01}.

In order to count  the number of blue and red supergiants in our lists, 
we had to apply corrections based  on our completeness tests.  
We used the completeness
values  described in Section 2.3 of Paper~I, which we summarized in  
its Table~3.
We   calculated the completeness values for two 
intervals of $\log(T_{\mathrm{eff}}) $  along each evolutionary track: 
from the end of the main-sequence to $\log(T_{\mathrm{eff}}) = 3.99 $ 
and for the interval  $\log(T_{\mathrm{eff}})  \le  3.70 $. For each of the points 
that define the evolutionary tracks  we interpolated linearly  in the completeness tables 
to calculate their completeness value. We also computed a weight defined as the
 difference between the age of this point on the evolutionary track, and the age 
 of the same point in the immediately older evolutionary track.  The adopted 
 completeness value is the weighted mean in each interval. 

Counting the blue supergiants in our sample presented  no problem; we simply 
calculated the number of stars within each mass range provided by the evolutionary 
tracks. However, if we compare the distribution of red supergiants in the \hrd ~to
that of the various stellar evolutionary models, we find that none of the models
produce RSGs as cool and luminous as what is actually observed. This fact 
was noted by  \cite{Massey03}, and \cite{MassOlse03}. They 
found the same problem while trying
to account for the RSG content in the Magellanic Clouds.
In their work, the sample of RSGs includes both stars with known spectral type and
a group of objects with known photometry but no spectroscopy.
They plot those supergiants in    [$\log(T_{\mathrm{eff}}) $, $M_{\mathrm{bol}}$] planes
and overplot  several sets of evolutionary tracks. They use a new calibration 
between spectral type and  $\log(T_{\mathrm{eff}}) $ to place stars with
known spectral type. For the rest of the stars, they use the intrinsic color $(V-R)_0$.
Whatever the method to place the stars in the diagram, we can argue 
that there is no significant difference.
Their Figure 5(a)  for the LMC RSGs is quite similar to our Figure~8 in Paper~I.
It is important to note that our method is strictly photometric, because \cho\
fits the best known SEDs from \cite{Kuru04} to a set of observed  photometric colors.

In order to count the
RSGs in our sample we had to artificially extend   the evolutionary 
tracks at constant  values of   $M_{\mathrm{bol}}$, and used the region
marked by a small rectangle located between the 15 and 25  $M_{\sun}$
tracks in Figures~8 and 9 in Paper~I  to the right of $\log(T_{\mathrm{eff}})  = 3.70 $.
After counting the red and blue supergiants in our lists we corrected these 
numbers for incompleteness. A summary of the results  of the ratio B/R is presented in  
Table~\ref{tbl01}. There, the columns labeled ``theory'' show the numbers 
derived from the ratio of time spent by a
star in each region and, therefore, assumes a constant star formation rate. 
The ``observation'' columns show the
numbers previously described in this paragraph.

\begin{figure}
\includegraphics[width=0.9\textwidth]{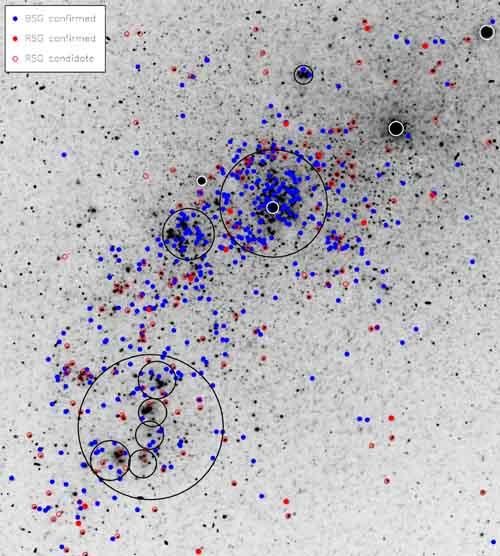}
\caption{Distribution of blue and red
supergiants on an F814W  mosaic of NGC~4214.
The field shown is the same as in Figure~\ref{fig01}.
The filled circles represent the confirmed supergiants, both blue and red that
we used to calculate the B/R ratio.
With open red circles we represent the group of stars  that follow 
the criteria
$\log(T_{\mathrm{eff}})  \leqslant 3.70 $ and  $M_\mathrm{bol} \leqslant  -6.0$.
and which might  be RSGs,  AGBs/ bright RSGs.
The cluster apertures are also drawn for reference.
}
 \label{fig02}
\end{figure}

Figure~\ref{fig02} shows  the distribution of blue and red
supergiants on an F814W  mosaic of NGC~4214.
The filled circles represent the confirmed supergiants, both blue and red that
we used to calculate the B/R ratio.
With open red circles we represent the group of stars  that follow the criteria
$\log(T_{\mathrm{eff}})  \leqslant 3.70 $ and  $M_{\mathrm{bol}}  \leqslant  -6.0$.
The latter objects may be either RSGs, AGB stars, or even bright red giants that 
have ``leaked'' into higher--mass
regions in the H--R diagram due to observational uncertainties.

Despite the   difficulty in counting RSG because the
computed grids do not produce RSGs as cool  and luminous as what 
we observe, we find a good agreement between our data and the non--rotating 
models with low metallicity $(Z=0.004).  $
These models predict a B/R value of 24 in the $ 15-20 \, M_{\sun}$ mass--range, and 
we measure $34 \pm 10$. Some of the stars below the  15 $M_{\sun}$ track
may be  RSGs, so we are  underestimating the  real number of 
RSGs.
In the mass--range $ 20-25 \, M_{\sun}$ the
theoretical prediction is 47 and we obtain $46 \pm 23$.
In both cases, our results agree with  the theoretical ones  within Poisson errors. 


There are two caveats regarding this result that should be mentioned. 
First, the small quantity of confirmed 
RSGs in our sample in NGC~4214 imply that stochastic effects due 
to small--number statistics may be present.
However, given the large differences in the theoretical ratios 
shown in Table~\ref{tbl01}, this effect is likely to
be unimportant. The second caveat is related to the conversion from 
observed colors to effective temperatures and
bolometric magnitudes. \cite{Leveetal05} present a new effective temperature scale for Galactic 
RSGs  by fitting MARCS stellar atmosphere models \citep{Gustetal75, Plezetal92} which include an 
improved treatment of molecular opacity to 74 Galactic $(Z=0.020)  $ 
 RSGs of known distance. They compare their location on the  
[$\log(T_{\mathrm{eff}}) $, $M_{\mathrm{bol}}$] plane with theoretical 
evolutionary models from \cite{MaedMeyn03} and find a much better agreement
between theory and observation.
Their main result is that RSGs appear to be warmer than previously thought.
This effect shifts the stars to the left in the diagrams making them coincide with the end of the tracks. 
It would be interesting to compare the change of temperature
from the fitting of observed optical colors using \cho\ or a similar code 
from Kurucz to MARCS atmospheres
to verify if it accounts for the $350-400$ K discrepancy detected in our data. 
More importantly, such a change should
also shift the stars downwards in the H--R diagram, since a higher
 temperature implies a lower
bolometric correction.

At the time in which we performed our fits, \cho\ did not
have the capability to fit  stellar models other than Kurucz or Lejeune. 
It is in our future plans to reanalyze this data 
using the MARCS stellar atmospheres models.

\section{Initial mass function}

\subsection{The IMF of the resolved population}
A lot of effort has been put into obtaining the IMF of the Milky Way and the Magellanic Clouds
 in the past decades.
Star counts  in clusters/associations in those local galaxies reveal an IMF with a slope close to Salpeter
$(\gamma = -2.35) $ above $\sim 1 M_{\odot}$  \citep{Scha03} when  a power law of the
form $\frac{dN}{dm}  = A \, m^{\gamma} $ is used.
\cite{Masse98} concludes that there is no difference in IMF slopes found between the Milky Way, the LMC
and SMC. He claims that the weighted average of the IMF slopes of the MW is $\gamma = -2.1\pm0.1$;
and that of the MCs is $\gamma = -2.3\pm0.1$. This shows that 
metallicity does not affect the IMF
slopes of massive stars.

It has long been known that some very massive stars are not currently
found in clusters or  associations, but they seem to be part of the field \citep{Masse98}.
Studies of the MCs showed that very massive stars can be found in highly
isolated regions, and that  they are the result of small star--forming events. 
An IMF study of these field objects led to surprising results: the actual IMF slope
is quite steep, with  $\gamma \sim -5$  \citep{Massetal95b}. This trend is found 
in the MCs and the Milky Way as well, and is easily detected in the spectral
type distribution \citep{Bergh04}.
It is currently debated whether the difference in slopes is due to
differences between in situ formation styles or to
a majority of field O stars being runaways \citep{deWietal05}.

What do we know about the stellar IMF above $\sim 1 M_{\odot}$ for galaxies beyond the MCs?
There are several problems  involved in the determination of IMFs in extragalactic systems. 
Only a few  galaxies are close enough for star counts to be carried out with any reliability, 
and even these are so distant that only the very brightest part of the luminosity function 
can usually be sampled. In addition, there are practical problems with star counts in 
external galaxies, including crowding, incompleteness, and corrections for foreground stars. 
Among other studies,
\cite{Veltetal04} studied the IMF of M~31 and derived a slope of $\gamma = -2.59\pm0.09$; 
\cite{Jameetal04} 
found an IMF slope of  $\gamma  = -2.37 \pm  0.16$ for the ionizing cluster of NGC~588 
in the outskirts of the nearby galaxy M~33; \cite{Annietal03} analyzed the star formation history
of NGC~1705 and inferred  an IMF slope close to Salpeter.

The  most direct and reliable method of obtaining the IMF of a certain
stellar population is based on counts of stars as a function
of their luminosity/mass. However, there are other indirect methods
which do not employ star counts, but still yield some information on the IMF \citep{Scalo86}.
One of these is the method of population synthesis, which attempts to match the
observed galaxy colors, spectrum or line strengths by finding the best mixture
of stars of various spectral types and luminosity classes. The main problems that arise
with this method are the uniqueness of the solution and the types of
astrophysical constrains that need to be imposed on the data.

\cite{Chanetal05}  apply this method to estimate the IMF slope in the field stars of NGC~4214.
They  assume that  the faint intra--cluster light in NGC~4214 corresponds to the 
field stars in the galaxy. They 
compared the spectroscopic signature with \stb\ evolutionary synthesis models
and found a lack of   strong O--star wind features in the spectra, which led them to conclude that
the field light originates primarily  from a different stellar population, and not from scattering 
of UV photons leaking out of the massive clusters.
Fitting IMF slopes in continuous--formation \stb\ models they infer that the best value 
for NGC~4214 would be   $\gamma =  -3.5  $.
Their work also provides similar values of the IMF slope for other local starburst galaxies
such as NGC~1741, NGC~3310, and NGC~5996.

Comparing   low resolution spectra taken 
 with the International Ultraviolet Explorer (IUE) 
in the 1100--3200 \AA\ range with the predictions of 
evolutionary population synthesis models, \cite{MasHKunt99} 
obtain $ \gamma = -3.0$ 
for the sum of field and clusters in NGC~4214.

\subsection{Method and sample selection}
Determining the IMF of a stellar
population with mixed ages is a difficult problem.
Stellar masses cannot be weighed directly in most instances, so the mass
has to be deduced indirectly by measuring the star's luminosity  and evolutionary state.
We followed the photometric/spectroscopic method provided by  \cite{Lequ79}: 
we estimated bolometric magnitudes 
(\mbol), and effective temperatures (\teff) for each star in the sample, as we 
explain in Paper~I.
 The present--day mass function was  
estimated by counting the number of stars in the \hrd  ~between 
theoretical evolutionary tracks computed
for models with different masses. 
In the case of a coeval star--formation region, the obtained function
{\em  is}  the IMF  modified by evolution at the high--mass end. 
For a continuous
star formation region, the IMF followed by division of the number of objects 
in each mass range
by the main sequence lifetime $(\tau_{MS})$ of the corresponding mass.

For the IMF determination, we used the theoretical evolutionary tracks from   \cite{LeSc01} 
in the mass range 5  to 120 $M_{\odot}$.
 These  theoretical models    provide 51 values of \teff, $L$ and age for 
each evolutionary track for a given metallicity. 
The first 11 points of each track define their main--sequence sections.
A line connecting the first point of each track defines the 
zero age main--sequence (ZAMS). 
This line is the left boundary of our main--sequence. A line connecting the 
11th point of each track defines the right boundary of the main--sequence.

Here we present a detailed analysis of the IMF of NGC~4214 using two key
assumptions: (i) we selected only main--sequence stars from our sample, and 
(ii) we assumed  that the  IMF has remained constant
  as a function of time.
Figure~2 in our Paper~I shows 
three regions of strong  \ha\  emission, which can be matched to active recent star--forming regions. 
These regions are I--A, I--B and II. 
In   {\sc\footnotesize  LIST336}  we found a total of  3061  objects located along  the main--sequence band.  
We considered  three lists of masses to calculate the IMF: (1) {\sc\footnotesize  LIST1}: a list including
all objects  (3061 stars), (2)  {\sc\footnotesize  LIST2}: a list with objects in regions  
I--A, I--B and II (active star--forming regions)
 (900 stars), and (3)  {\sc\footnotesize  LIST3}: a list with all objects  except those of  I--A, I--B and II  (2161 stars.)
We assumed all binary and higher--order stellar systems are resolved into individual stars. See
Section 4.3.3  for a detailed discussion on multiplicity.

We started by determining the stellar masses and their errors
of all objects in the lists
using the estimated values of \mbol ~and \teff    ~obtained with \cho\ as described in Paper~I.
We plotted a \hrd  ~[$\log$ \teff , \mbol ] on which we superimposed
theoretical evolutionary tracks from   \cite{LeSc01} in the mass range $5-120 M_{\odot}$.
We then isolated  the main--sequence sections of those tracks and we obtained interpolated mass values 
between the tracks using splines. We built an  IDL code that triangulates the whole grid and interpolates 
the data to obtain a finely spaced grid of masses throughout the whole \hrd. This procedure gave us a 
two variable function $M = f($~\teff,~\mbol~) which assigns a mass $(M)$ value 
from the input values \mbol ~and \teff. 
The individual stellar masses $(M_i)$ followed directly from this function.
To obtain the uncertainties $(\sigma_{M_i})$  for the individual masses $(M_i)$, 
we created a randomly distributed
set of 10000 points around the mean values of \teff~   and \mbol~   for 
each star and  inferred
 their individual masses. Those points were distributed using 
 a bidimensional gaussian distribution according to the parameters provided
 by the output of {\sc\footnotesize  CHORIZOS}.
The standard deviation of these masses gave us  a reliable value of the error 
 $(\sigma_{M_i})$  in the mass  $(M_i)$ of each star.  
 How dependent are our results on the evolutionary models used for the mass tracks in the \hrd ~? 
We found that the locus  of an  evolutionary track   of a given mass on the \hrd, changes with initial
conditions  such as metallicity, mass--loss, rotation, etc. Most of them show loops which 
give them a  complex shape.
However, these changes are minor and may be neglected when we consider the uncertainty 
in the mass determination for the objects in our lists, especially since 
we are considering objects that lie within the limits of the main sequence where
those changes are insignificant. 
The covariance ellipse for each single
object, usually spans a wide range of evolutionary tracks (due to the intrinsic errors
in the determination of  \teff\ and \mbol\ from \cho.) This would have not changed if
we had chosen a different set of theoretical models. 
It is important to note that the most massive objects in our lists have 
smaller relative errors
than the less massive ones. 
How does that translate into the determination of the error in the mass of each star? 
Even though the most massive objects have smaller errors in  \teff\ and \mbol, 
their errors in mass 
 may still be large 
because their covariance ellipses extend over  a wider range of masses.
On the other hand, objects with lower masses  have bigger errors in    \teff\ and \mbol, but
their location in the \hrd ~is such that their ellipses span a narrower  range of 
evolutionary tracks.

Regions  I--A, I--B and II  are young clusters with recent bursts of star formation.
They all have different ages as we will show in Section~5. For these regions, we 
considered that 
the present--day mass function {\em is} the IMF.
The rest of the galaxy is  composed of a large number of star--forming regions
of different ages, and a fairly good approximation would be to 
consider it as a region of  continuous star formation.  For this case we corrected the star counts for 
the main--sequence life--time $\tau_{MS}$ at each  particular mass, using values from Table~4 in Paper~I.

\subsection{Correction of systematic effects}

In order to determine the IMF of our sample of stars, we 
analyzed four sources of systematic effects, as pointed out by  \cite{Maizetal05}:
(1) data incompleteness,
(2) appropriate  bin selection,  (3) unresolved objects in our sample,
and  (4)  mass diffusion.  

\subsubsection{Data incompleteness}

In the derivation of an IMF, a quantitative 
evaluation of completeness
of the photometric data is required.  
To obtain the completeness values along the main--sequence section of each evolutionary track,
we calculated its value on each point that defines the main--sequence using
the results from Section 2.3  in Paper~I; we also 
computed a weight defined as the difference between the age of this point on the evolutionary 
track, and the age of the corresponding point in the  nearest more massive   evolutionary track. 
The adopted  completeness value is the weighted mean in each interval. All values are
summarized in  Table~4 in Paper~I.
We determined the main--sequence lifetime of each available track by interpolating among the ages 
provided by the models.

\subsubsection{Bin selection}

To calculate the slope of the IMF we followed the technique suggested
by      \cite{MaizUbed05}  
and used a weighted least square method
to fit  a power law $\frac{dN}{dm}  = A \cdot m^{\gamma} $
using   5 bins of  variable size, so that the number of stars in
each bin is approximately constant. This method guarantees that  the binning biases
will be minimum 
and that the uncertainty estimates are correct.
For each   bin, we used the weight $w_i$ derived from
a binomial distribution as  suggested by  \cite{MaizUbed05}.

\subsubsection{Unresolved objects}

In distant stellar systems like the MCs or NGC~4214, we encounter a serious problem: namely, 
crowding of stellar images which makes the IMF determination somewhat complicated.
At the distance of NGC~4214 (2.94 Mpc), 1 arcsec corresponds to $\approx$14 pc. We therefore
expect some of 
the regions in the galaxy to be  crowded from the observational point of view.
We may find  unresolved objects in our sample, i.e., merged stellar images, irrespective of
their nature as a physical system or a chance coincidence.
The reliability of the highest mass stars is questionable because of  stochastic effects due to  
small number statistics  and evolutionary effects. Some stellar evolution has 
certainly taken place for these 
stars and some of them may fall out of the main--sequence band.

The effect of unresolved objects in a sample used to determine a stellar IMF has been 
already analyzed by several authors.  
\cite{SaRi91} studied the effect of actual binaries
in the mass range $2-14 ~M_{\odot}$ by performing several Monte-Carlo experiments
using initial  values $\gamma = -3.5, -2.5,$ and $ -1.5$ and several binary fractions.
They come to the conclusion that the general effect 
is a flattening of the mass function.
With an intrinsic slope of $\gamma = -3.5$ there is hardly any influence
on the MF slope; and for $\gamma = -2.5$    the  MF slope undergoes a 
flattening of 0.34 if the binary fraction increases to 50\%.
They made the
assumption that stellar masses are distributed randomly among the components.
\cite{Krou91} show that unresolved binary stars have a significant effect on any 
photometrically determined luminosity function. Their research leads to the conclusion
that when binaries are not taken into account, there is an underestimate on the 
number of low mass stars, leading to a flattening of the IMF slope.
Given the large multiplicity fractions 
observed for OB stars, this effect must be significant for intermediate-- and high--mass stars 
\citep{Maizetal05}. However, since the binary fraction is unknown  in our sample,
it is not possible to correct  the estimated  values of the IMF slope for the presence of binaries. 
We used $M_{up} = 100  M_{\odot}   $ as the high--mass end of the integration which  allowed us to 
discard some  unresolved objects in our sample, but 
this effect is difficult to solve for a galaxy located at the distance of 
NGC~4214 because, even with 
the high--resolution images provided by HST, we still observe 
multiple systems. 

One additional  source of multiple objects  is  the blends of stars caused by 
chance alignments. The greater the distance to the object of study,
the more likely one is prone to encounter  these kind of objects.
Our analysis of multiple systems 
leads us to conclude that
the real value of the slope would be steeper than the 
value that we obtain with our study.

For the low--mass end of the integration, we used several  mass values  in the range 
$6, 6.5, ..., 25  M_{\odot}   $. The main problem regarding the low--mass end is the completeness
of our data, because the observational sensitivity limit is reached here.
We corrected the number of counts for incompleteness using the values given in  Table~4 in Paper~I,
as described above. The IMF slope values that we present were calculated for 
 $M_{low} = 20  M_{\odot} $  where the completeness is higher than 85 \%.

\subsubsection{Mass diffusion }

Using \cho\ we  translated uncertainties in the measured magnitudes and colors
into uncertainties in temperature (or bolometric magnitudes) and 
luminosity as we explained in Paper~I.  
We then  used the results from Section~4.2  to obtain the 
uncertainties in mass.
For our stellar mass range, the IMF has a negative slope, meaning
that there are more low--mass stars than high--mass ones.
Therefore, if 
we measure a star to have magnitude $m'$, there should be a higher probability 
that its real magnitude $m$ is dimmer than $m'$ than it is  brighter (i.e. there 
are more dim stars disguising as bright ones at a given measured magnitude 
than bright stars disguising as dim ones).
\cite{Maizetal05}  study this  artifact that appears when dealing with real data.
This  effect basically
smoothes the IMF slope by shifting dim objects (low--mass stars)
to where bright objects  (high--mass stars) are, and it is basically
due to the effect of uncertainties in temperature and luminosity of individual
stars.
To take care of this effect, we proceeded in the following way: First, a
function  $(\sigma_{M}(M))$ was computed from the values for the individual 
stars  $(M_i , \sigma_{M_i})$
by fitting a third degree polynomial. Then,  we used a random number generator to produce 
50 lists
of masses in the range $  [10,200]  M_{\odot}$  for different  input values of  the IMF slope
which we called $\gamma_{real}$. Those values are in the range $-1.2 $ to $-4.0$ with step of 0.1.
We assigned  errors randomly  to the masses in our artificial lists to smooth the values
using   the function   $\sigma_{M}(M)$, and 
 simulated the incompleteness of the data using the values given in  Table~4 in Paper~I.
We then  fitted these lists with a power law using   5 bins of  variable size, so that the 
number of stars in
each bin is approximately constant. For the integration, we  used the same mass
limits   that
we had used for the real data.
This process allowed  us to fit  a polynomial to $\gamma_{fit} - \gamma_{real}$
as a function of $\gamma_{fit}$  which is the correction that needs to be applied
in order to obtain $ \gamma_{real}$ from the estimated $\gamma_{fit} $.

\subsection{IMF  Results}

The result of the process described in Sections 4.2 and 4.3
 is shown in Tables~\ref{tbl02} and ~\ref{tbl03}.
For each set of stars, we give the computed values 
 $\gamma_{fit}$, and $ \gamma_{real}$ that we obtained for six values of 
 the low mass   ($M_{low}$ ) used in the integration. We also present 
 the uncertainties  ($\sigma$) in $\gamma_{real}$  derived from the $\chi ^2$  fit 
as well as the number $(N)$ of stars used for each fit. 
Table~\ref{tbl02} gives the results for    
a continuous star--formation scenario.
Table~\ref{tbl03}  gives the results for    
a burst star--formation scenario.
We adopt  $ \gamma_{real} = -2.83 \pm 0.07$  as a 
representative value of the IMF slope of NGC~4214,  which we calculated
using  $M_{low} = 20 \, M_{\odot}   $ and $ M_{up} = 100 \, M_{\odot}$.
We used IDL's CURVEFIT  to fit the power law and the $\chi^2$ value
provided by the fit in the [20, 100] $M_{\odot}$ mass range is 1.94.

%
%
%
Since the correction for unresolved objects is uncertain  (given the large distance to
NGC~4214), and knowing that unresolved objects produce a systematic 
flattening of the IMF slope, we conclude that 
the real value of the slope calculated for {\sc\footnotesize  LIST1}
 would be steeper than $ \gamma= -2.83$.
The burst star--formation model may be only applied to the stars included in  
regions I--A, I--B and II ({\sc\footnotesize  LIST2}),
because these are compact star--forming regions produced as a single burst.
%
Note that the correction supplied by the  polynomial is a decreasing  function of 
$M_{low}$. This correction is greater for low values of $M_{low}$  where the incompleteness of the
data plays a strong role. 
When considering objects from {\sc\footnotesize  LIST2}, one has to face two problems:
(1)  the  uncertainty 
in the  correction for unresolved objects as described above for  {\sc\footnotesize  LIST1},
and (2) the fact that these regions are composed of a mixture of stars with ages in
the range $0-5$ Myr.  This means that some of the
most  massive stars must have evolved and disappear
in the form of supernovae  emptying the high--mass bins and producing a steeper slope.
We studied this problem by considering stars in the following
mass ranges (in units of solar masses): [20,  40],  [20,  100],  and  [40,  100].
The values of the IMF slope are  $ \gamma_{real} = -4.41 \pm 0.28$,  $ \gamma_{real} = -3.56 \pm 0.13$,
and  $ \gamma_{real} = -2.97 \pm 0.37$  respectively.
These different values could be explained by claiming that
the fraction of multiple and blended systems is larger in the   [40,  100] range than in the 
 [20,  100].
This makes the slope to appear shallower than when using  $ M_{up} = 40  M_{\odot}$. 
In summary,  $ \gamma_{real} = -3.56$ is a lower value for the IMF slope
when using the burst star--forming model.

\cite{Chanetal05}  estimate the value $ \gamma = -3.5$
for the field stars in NGC~4214  by matching integrated SEDs
to STIS spectra, 
and using $M_{low} = 1  M_{\odot} $ and 
$M_{up} = 100  M_{\odot}$. Their extracted field regions include
star clusters, and hence some additional massive stars that do not belong to the field.
This implies that their slope value represents a lower limit to the field IMF slope.
As a consequence of this, they argue that the field is much less likely 
to produce massive stars than the cluster environment.
Using a similar approach, but with $IUE$ spectra, 
\cite{MasHKunt99}  could constrain the IMF slope  of NGC~4214 to
$ \gamma = -3.0$. Their spectra clearly includes stars in clusters as well as field stars.
We expect the ``real'' IMF slope to be shallower than Chandar's and 
steeper than our calculated value ($ \gamma= -2.83$).

\section{Analysis of extended objects}

In Paper~I we described the process that we used to obtain aperture photometry 
of 13  stellar clusters ( I--As, I--Es, IIIs, IVs, I--A, I--B,
 I--Ds, II, II--A, II--B, II--C, II--D, and II--E )
in NGC~4214. The magnitudes are listed in  Tables~3 and 4
of that paper. We also   explained 
how we translated the observed magnitudes and photometric colors
into clusters  physical properties using \cho\ ~\citep{Maiz04}. 
We based our  work on clusters on   
\stb\ \citep{Leitetal99}  models  of integrated stellar populations.
Here we analyze the results.

\subsection{Unresolved clusters}
\subsubsection{Cluster I--As}

We start our analysis with cluster I--As. 
\cho\ was executed using the clusters
magnitudes from Table~6 in Paper~I, and  
leaving  all three parameters [$\log({\rm age}) $, \ecc ,  $R_{5495}$] 
unconstrained, which  gave us a well defined value 
of the clusters age: $\log(\rm {age/yr})=6.60     \pm 0.06 $, and a low value for the extinction 
$E(4405-5495) =  0.07 \pm 0.02$ mag.
The reddening vs. age likelihood contour plot (not shown) clearly indicates
that  there is only one solution compatible with the available photometry.
However, the low value obtained for the reddening 
implies a degeneracy in the \rv ~values, since the effect of  extinction  becomes
nearly  independent 
of the choice of extinction law   when \ecc $\ll 1.0$ mag.  This means that even though
we are able to restrict the age and extinction of this cluster, 
the appropiate value of \rv ~remains undefined using our set
of observed magnitudes.

\begin{figure}
\includegraphics[width=0.9\textwidth]{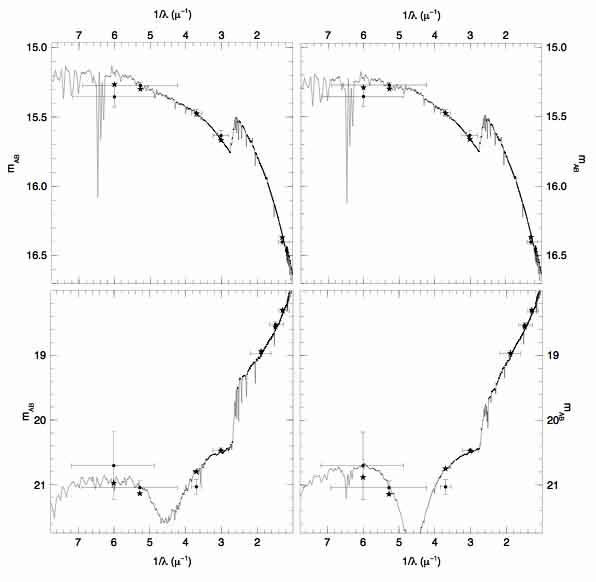}

\caption{SEDs for  the best fits of clusters I--As and I--Es, based on \stb\ models
of integrated stellar populations. The photometry is shown by the symbols
with error bars (vertical ones for uncertainties and horizontal ones for 
the approximate length coverage of each filter.) Star symbols indicate
the calculated magnitude of the model SED for each filter.
[Top left] Best fit SED for cluster I--As using an LMC2--like extinction law.
[Top right] Best fit SED for cluster I--As using an SMC--like extinction law.
[Bottom left] Best fit SED for cluster I--Es for the old (189 Myr) solution.
[Bottom right] Best fit SED for cluster I--Es for the young (7 Myr) solution.  }
\label{fig03}
\end{figure}

\begin{figure}
\includegraphics[width=0.9\textwidth]{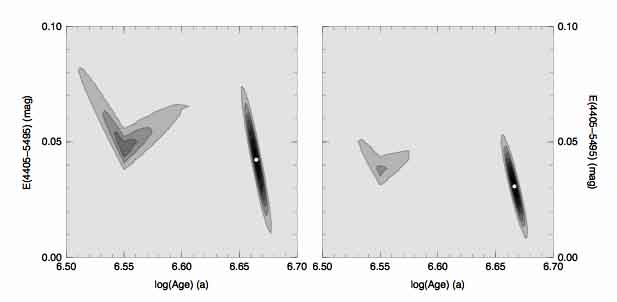}

\caption{  Reddening vs. age likelihood  contour plots for cluster I--As provided as output by \cho\ 
and obtained using \stb\ models.
[Left] Contour plot obtained using the LMC2 extinction law. 
[Right] Contour plot obtained using the SMC extinction law. 
Note that the age of this cluster is in the range 3.0--4.5 Myr, and that it 
is characterized by a very small extinction. The white circle marks the most
likely value (mode).
  }
\label{fig04}
\end{figure}

Running the code with the same set of magnitudes, but constraining the extinction law
to a MC--type (average LMC, LMC2, or SMC) we obtained similar values, 
as shown in  Table~\ref{tbl04}.
Since we are fitting two parameters
and using four colors (derived from five magnitudes), the problem has 2 degrees of
freedom. 
The best fit, which is certainly degenerate,  appears to be 
the one obtained using the SMC extinction law,
according to the $\chi^2$ reduced value. 
Figure~\ref{fig03}    shows the SEDs that best agree with  the data for the LMC2
and SMC  fits. In Figure~\ref{fig04}  we   show the reddening vs. age likelihood contour
plots using the LMC2 and SMC extinction laws. Both plots show that 
the age of this cluster spans  the range  $6.50 \le   \log(\rm {age/yr}) \le  6.70 $.

\cho\ provides a time--dependent correction that can   
be applied to transform the current magnitude of a cluster to the one at zero age.
We used this quantity to derive  the  clusters zero age masses. For our
analysis, we used \stb\ models 
which   only include stars more massive than  $ 1 \, M_{\odot}.$ 
Since most of a cluster's mass is contained in the low--mass stellar population, we had to 
further correct these values  considering a realistic IMF that spans the mass 
range $0.1-100 \, M_{\odot}.$
All our mass estimates in this paper
were  therefore calculated using a Kroupa IMF \citep{Krou02}
which considers stars in the range $0.1-100 \, M_{\odot}.$

The derived mass of cluster I--As is $\approx 27\,000 M_{\odot}.$ 
Once the approximate mass
of the cluster was known, we used the \stb\ output to calculate the stellar content of a cluster
with  this mass,   an SMC--like  metallicity $(Z=0.004)$  and the  appropriate age.
The results of the evolutionary synthesis models are  shown in Table~\ref{tbl04}.
It is  known that cluster I--As is a SSC with a compact core and a strong massive halo,
very similar in structure to 30 Doradus in the LMC, as shown by 
\cite{Maiz01b}. This implies that aperture effects are 
very important when performing any kind of photometric measurements
for this cluster, and the derived quantities such as the mass of the cluster
and the stellar population will depend on the chosen aperture. 
 
Using an SMC extinction law, our models yield an age of $4.0 \pm 0.6$ Myr for 
cluster I--As from the optical--UV
photometry. This value agrees with the ones obtained by \citet{Leitetal96} 
using UV spectroscopy (4--5 Myr) and
by \citet{Maizetal98} using $W$(H$\beta$)  and the strength of the WR 4660 \AA\ blend (3--4 Myr).
The models that best fit our data yield the values\footnote{Note that the value of \ecc ~is 
slightly different from the one without restraining the extinction law
but that  both  are separated by only $\approx 1 \sigma.$} $A_V = 0.19  \pm 0.06,$ and  
$E(4405-5495) =  0.04  \pm 0.01 $ mag . This low extinction is expected, since I--As is located within a 
heart--shaped \ha\ cavity created by the kinetic energy input of the cluster into its surrounding medium
\citep{Maizetal99a,MacKetal00} and, therefore, one would expect little gas or dust in the line 
of sight. 
The quite large number of  massive stars formed here have apparently wiped 
out the dust particles from their surroundings, leaving free paths through which
the stellar continuum can emerge.
The  low extinction agrees with the value measured at this location by \citet{Maizetal98} 
from the nebular  ratio of \ha\ to \hb. 

From the detection of a broad emission feature from Wolf--Rayet stars (WRs), several authors   
\citep{Leitetal96,SargFili91,MasHKunt91a,Maizetal98}
indicate the presence of WR stars inside cluster I--As. 
\cite{Leitetal96} reexamined the results of \cite{SargFili91}.
Their  results were calculated from  ultraviolet 
spectra of cluster I--As  obtained with
the Faint Object Spectrograph onboard the HST in 1993. They used the $1  \farcs 0  $ 
circular aperture
and assumed a distance of 4.1 Mpc to NGC~4214.
They estimated that approximately 
15  WN and 15 WC  stars are present inside cluster I--As, keeping
in mind uncertainties of a factor of 2. These results translate to 5 WR stars 
using the improved distance of 2.94 Mpc.
\cite{Maizetal98}  study four regions with WR stars in NGC~4214, including cluster I--As. 
Using their  observed values 
of  the equivalent widths of the WR blend we can estimate 
60 WR stars inside an area of radius $1  \farcs 8 $ assuming  a distance of 
4.1 Mpc.  This result is equivalent to  3 WR stars
inside the aperture radius that we had used for cluster I--As at a distance
of 2.94 Mpc.
The \cho ~results  predict 4 WRs 
using the same metallicity $(Z=0.020)$, and 1 WR using  $(Z=0.004)$,
which is consistent with the observations.
It is important to note that WR stars may be present both in the nucleus of the cluster, 
and  they may also 
be part of the massive halo. We want to emphasize that aperture effects are of great significance 
in dealing with this cluster. This  makes it hard to compare different works performed with 
different techniques; however, our results are within Poisson errors of previous published values.

\subsubsection{Cluster I--Es}
\begin{figure}              
\includegraphics[width=0.9\textwidth]{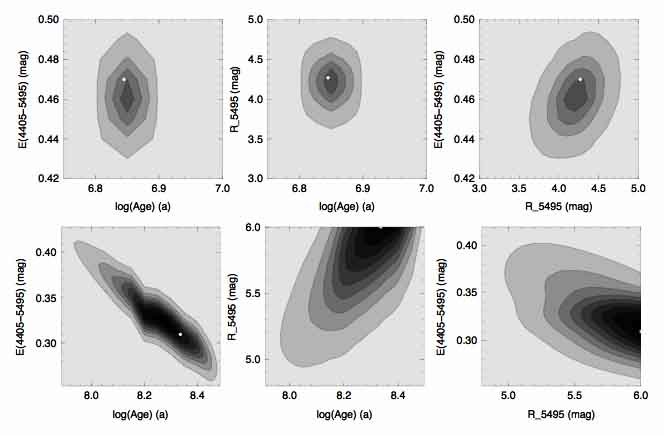}

\caption{ Likelihood contour plots for cluster I--Es provided as output by \cho\ and obtained using \stb\ models.
The upper row corresponds to the young solution and the lower row to the old one.
[Left] Reddening vs. age plot. [Center] Extinction law vs. age. [Right] Extinction
law vs. reddening. Age is expressed in years. The white circle marks the most
likely value (mode). Note that the mode in 3D parameter space (shown here) 
does not necessarily coincide with the mode
in 2D parameter space. 
}
\label{fig05}
\end{figure}

We executed   \cho ~for cluster I--Es using  seven observed magnitudes (six colors)
from Table~6 in Paper~I and  leaving all parameters unconstrained.
The likelihood map produced by \cho\  shows two solutions in the [$\log({\rm age})$, \ecc] plane,  
a young one around $\log({\rm age/yr}) \approx 6.85$, and an old one around $\log({\rm age/yr}) \approx 8.25$.
We reran the code to isolate these two solutions using those age intervals,
and the properties of both solutions are shown in Table~\ref{tbl04}. 
The old solution (age of $189\pm53$  Myr)
is the one that has the highest probability, but the young solution cannot be immediately rejected.
The likelihood contour plots produced by \cho\ for this cluster are shown
 in Figure~\ref{fig05}   for both solutions. The young solution shows 
very well defined ellipses in the plots. The old solution yields a high value of
\rv ~which is clearly visible in Figure~\ref{fig05}  (lower row, center and right.)
The SED for both models  are given in Figure~\ref{fig03}.
Both solutions show that cluster I--Es  is older that cluster I--As
which  agrees with 
\cite{MacKetal00} who indicate that 
 the continuum colors of I--Es  are significantly redder than for I--As,
 indicating a greater age. 
 We executed \stb\   to estimate 
the stellar content of a cluster with mass, age and metallicity appropiate 
for both  solutions. The young solution  ($\approx 6\,000 M_{\odot}$ )  predicts the
 presence of 1 red supergiant
and 2 blue  supergiants, while the old solution ($\approx 129\,000 M_{\odot}$ )  predicts
9 red supergiants and zero blue supergiants.

\begin{figure}
\includegraphics[width=0.9\textwidth]{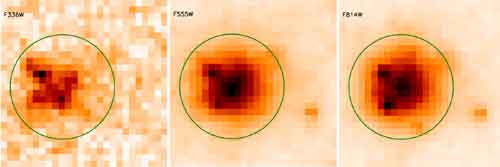}
\caption{Images of cluster I--Es obtained with HST/WFPC2 using filters 
F336W, F555W, and F814W. The images have been 
resampled in order to use the same linear scale in all cases, with the field
 sizes being $15.9  \times 15.9$ pc$^2$.
The orientation in each case is north pointing up and east pointing to the left.
Special attention must be drawn to the two objects located 
off center from  the cluster. See the text for a complete discussion. }
 \label{fig06}
\end{figure}

Trying to discern between the two possible solutions leads us to give
a word of caution about the predictions of stellar synthesis models: in general, theoretical
models have intrinsic uncertainties, especially when dealing with fast--evolving phases such as the RSG one.
Furthermore, a more subtle effect is caused by incomplete sampling.
Most current synthesis codes predict the average values of observables for an {\em infinitely large population}
of stars, which samples a given IMF completely. \cite{CervVall03} show that for an integrated property which
originates from an effective number of stars $N$, there is
a critical value of $N\approx 10$ below which the results of the codes must be taken with caution, since they 
can be biased and underestimate the actual dispersion of the observables. 
Cluster I--Es is an example of this sampling problem, since its F814W magnitude is dominated by the presence of
a few RSGs (both the old and young solutions predict this). Therefore, the integrated colors are not enough to 
differentiate between the two solutions, since stochastic effects should be larger than evolutionary ones.
Is there a way to overcome this issue? One could do it by looking at the resolved stellar population of the
cluster, which is what we attempt in Figure~\ref{fig06} by
representing cluster I--Es in detail in three filters: F336W,
F555W, and F814W. A close inspection of these plots reveals that the cluster
has some  non--radial  substructure.
These  images show two objects that blend with the structure of cluster I--Es,
which are located inside the aperture radius that we used for the photometry
marked in green in Figure~\ref{fig06}.
The  object located towards the north   is clearly visible in  three filters (and is likely to be a 
BSG), while the other is not detectable in filter F336W (and is likely to be a RSG).
The presence of a single bright red object favors the young solution over the old one, given that the first one
predicts 1 RSG and the second one 9 of them. On the other hand, the bright red object could simply be a RSG 
unassociated with the cluster that happens to lie in the same direction (a look at Figure~\ref{fig02} shows that
this possibility is not unlikely). Therefore, we have to conclude that the  available data does not allow us to 
discern between the two solutions.

Regarding the extinction that affects I--Es, both solutions (young and old) yield values much higher than for I--As.
This is consistent with the \ha/\hb\ results of \citet{Maizetal98} and \citet{Maiz00} and with the existence of a
small peak in the CO distribution near the position of the cluster \citep{Waltetal01}. Also, it is interesting to
note that the $R_{5495}$ obtained from either solution is higher than the standard 3.1. However, given the
uncertainty in the characteristics of this cluster deduced from its integrated colors caused by its relatively low
number of stars, we should regard this measurement of $R_{5495}$ as rather uncertain.
 
\subsubsection{Cluster IIIs}

\begin{figure}              
\includegraphics[width=0.9\textwidth]{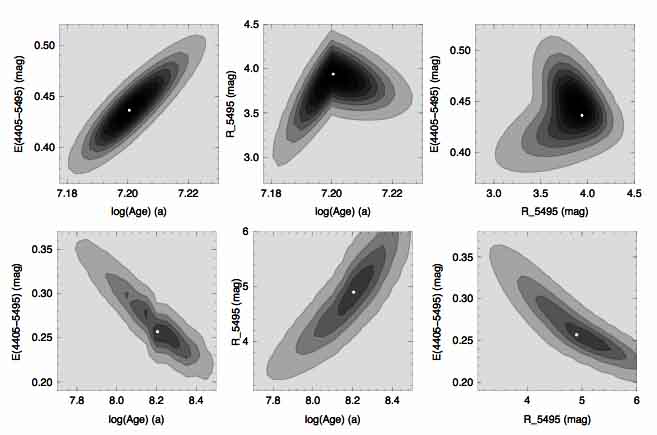}

\caption{ Likelihood contour plots for cluster IIIs provided as output by \cho\ and obtained using \stb\ models.
The upper row corresponds to the young solution and the  lower row to the old one.
[left] Reddening vs. age plot. [Center] Extinction law vs. age. [Right] Extinction
law vs. reddening. Age is expressed in years. The white circle marks the most
likely value (mode). }
\label{fig07}
\end{figure}

\begin{figure}
\includegraphics[width=0.9\textwidth]{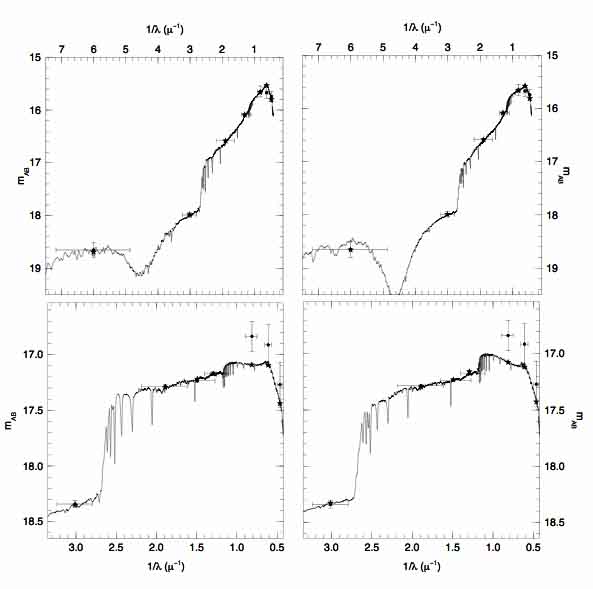}

\caption{ Same as Figure~\ref{fig03} for  clusters IIIs and IVs.
[Top left] Best fit SED for cluster IIIs for the old (167 Myr) solution.
[Top right] Best fit SED for cluster IIIs  for the young (16 Myr) solution.
[Bottom left] Best fit SED for cluster IVs for the old (150 Myr) solution.
[Bottom right]  Best fit SED for cluster IVs for the young (16 Myr) solution.
 }
\label{fig08}
\end{figure}

To analyze cluster IIIs, we used all seven
magnitudes   listed in Table~6 from Paper~I. We executed   \cho\ without limiting any  parameter 
and using F336W as the reference filter. This gave us 7 magnitudes, 6 colors and
3 free parameters.
We found an excellent fit of the spectrum with two possible solutions: a weak (high $\chi^2$) young solution
around $\log({\rm age/yr}) \approx 7.20$ and a stronger (low $\chi^2$) around $\log({\rm age/yr}) \approx 8.20$.
We reran the code twice to isolate these two solutions and 
the results for both models are presented in Table~\ref{tbl04}.
The likelihood contour plots for both solutions are presented in Figure~\ref{fig07}.
Both solutions are very good fits, with small values of $\chi^2$.
The slight difference between the two SEDs (Figure~\ref{fig08}) lies in 
the depth of the bump located
around  2175 \AA.This spectroscopic 
feature  is known as  the graphite bump.
Our study cannot disentangle between both solutions, since we do not have 
any measured magnitude in this part of the spectrum. However, the relative position of this cluster
in the galaxy suggests that  a low value of the extinction is to be expected.
Also, the image of the cluster appears to be smooth, with no salient features 
throughout its structure.
These two properties suggest that the old solution 
 (age of $168 \pm 61$  Myr, mass of  $(6 \pm 2) \times 10^5$  $M_{\odot}$ )
should be preferred over the young solution.

\subsubsection{Cluster IVs} 

\begin{figure}
\includegraphics[width=0.9\textwidth]{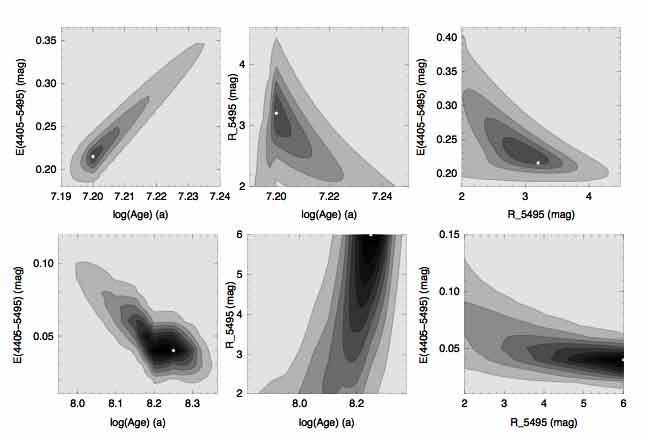}

\caption{ Likelihood contour plots for cluster IVs provided as output by \cho\ and obtained using \stb\ models.
The upper row corresponds to the young solution and the lower row to the old one.
[Left] Reddening vs. age plot. [Center] Extinction law vs. age. [Right] Extinction
law vs. reddening. Age is expressed in years. The white circle marks the most
likely value (mode).   }
\label{fig09}
\end{figure}

To analyze cluster IVs, we used all seven
magnitudes   listed in Table~6 from Paper~I. We executed   \cho\ 
without limiting any  parameter 
and using F555W as the reference filter.
There are two distinct  solutions compatible with our set 
of magnitudes:  a weak young solution
around $\log({\rm age/yr}) \approx 7.20$
and a more conspicuous old 
around $\log({\rm age/yr}) \approx 8.20$.
Figure~\ref{fig08} represents the best fit spectra for both solutions, and  Figure~\ref{fig09}
shows the probability contour plots for the young (upper row) and old (lower row) 
solutions. 
With our current magnitudes, it is not possible to decide between both models. 
However, for  cluster IVs we expect a low value of the extinction since it is located 
in a   region of the galaxy  with little gas and dust. Also, as in cluster IIIs, its smooth
appearence favors the old solution (age of $150 \pm 34$  Myr) over the young model.
As in cluster I--As, we find   large values of  \rv ~(spanning the range
$2.0  \leqslant R_{5495}  \leqslant 6.0$), indicating  that 
for a low value of the extinction, \rv ~is degenerated.
In order to disentangle this problem,
it would be useful to measure a magnitude near the right side of  the 
Balmer discontinuity  such as F439W (WFPC2 $B$).

In  Table~\ref{tbl04} we present the number of K+M stars of types I and II, obtained from 
\stb.
It is noticeable  that in   clusters IIIs and IVs, the young solution provides a much lower quantity 
of RSGs than the old solution. The larger number of RSGs would explain the smooth 
appearance observed in the images, and this is another fact that favors the older solution
for clusters IIIs, and IVs.
\cite{Larsetal04} estimate  $\sim (200 \pm 52 ) $ Myr as  the age of clusters IIIs and IVs,
which is consistent with our results.

\subsection{Resolved clusters}
\begin{figure}
\includegraphics[width=0.9\textwidth]{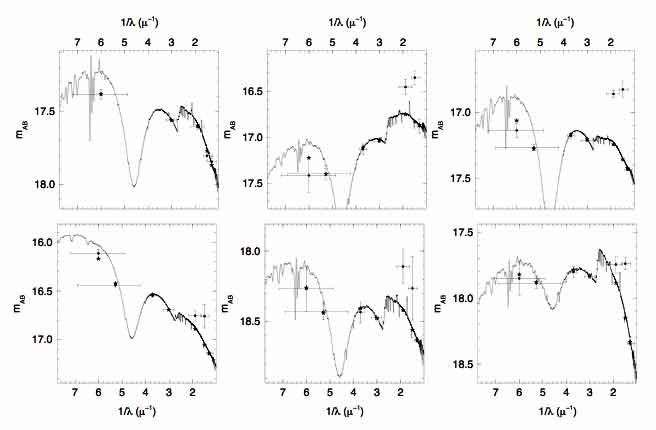}

\caption{ SEDs for  the best fits of clusters I--Ds [Top--left], II--A [Top--center],
II--B [Top--right],  II--C [Bottom--left], II--D [Bottom--center], and II--E [Bottom--right],   
based on \stb\ models
of integrated stellar populations. The photometry is shown by the symbols
with error bars  (vertical ones for uncertainties and horizontal ones for 
the approximate length coverage of each filter.) Star symbols indicate
the calculated magnitude of the model SED for each filter. 
Magnitudes in 
filters F555W and F702W are displayed in the spectrum of NGC~4214--IIABCDE, but
they were not considered during the execution of \cho\ because of possible
contamination.   }
\label{fig10}
\end{figure}

We used five  magnitudes from   Table~7 in Paper~I
to execute \cho ~for cluster I--Ds. We left 
 all parameters
[  $\log({\rm age}) $, $E(4405-5495)$, $R_{5495}$ ] unconstrained, which 
means that we had only one degree of freedom in this run.
Figure~\ref{fig10} shows the best fit spectrum to the measured data.
The contour plots  for this cluster are presented in 
Figure~\ref{fig11}  (left column.)
There appears to be two possible solutions, both of them very young, with a strong peak 
at   $\log({\rm age/yr}) \approx 6.5 $.  The output of the code yields
a mean age of  $2.6 \pm 1.5  $ Myr, and a mean  extinction  of 
$ E(4405-5495) =  0.28 \pm 0.05$ mag. The results are listed in Table~\ref{tbl05}.

The rest of the clusters which are part of the structure of  complex II,
were analyzed in a similar way. We used the set of magnitudes
 listed in  Table~7 from Paper~I  with the exception of magnitudes F555W and F702W
 for the reason explained below. We obtain fairly good fits 
 (highest $\chi^2 = 2.53 $ ) which we display in 
 Figure~\ref{fig10}. Note that the points that correspond to 
 magnitudes F555W and F702W are plotted, but they were not considered 
 during the  execution of \cho\ because those magnitudes are heavily contaminated by nebular emission. 

The equivalent width of the Balmer lines can be used to estimate the
age of a star--forming region \citep{CervMasH94}. The expected values for 
$W$(H$\alpha$) are $1000-2500$ \AA\ for very young clusters (age $ \lesssim 3$ Myr) 
and $500-1000$ \AA\ for ages in the range between 3 and 4 Myr.
\cite{MacKetal00}  estimated this parameter using three extinction corrections for all
the clusters in NGC~4214. 

The first execution for cluster II--A showed two possible solutions
compatible with our data. One in the range 
  $0.1  \leqslant   E(4405-5495)     \leqslant 0.26$ (in mag) and the other
in the range   $  0.26  \leqslant   E(4405-5495)   \leqslant 0.6$ (in mag). 
We reran the code to isolate the solutions individually, and 
here, we present   the diagrams corresponding 
to the second  case only.
\cite{MacKetal00} provide values of $W$(H$\alpha$) for all the clusters
within NGC~4214--II.  These are close to or higher than $1000$ \AA\ 
indicating a very young  age of this complex.  This fact led us to 
favor the solution with the highest $E(4405-5495)$.
With this method we can restrict the age of this cluster to 
$3.1 \pm 1.4  $ Myr. This means that there may be WR stars
within its structure. We also obtain a large value of \rv, compatible 
to the one  calculated for cluster II.

As for cluster II--B, the first execution showed two 
potential solutions with the same mean value of \rv.  
We isolated a young and an old solution.  Here we present 
the plots that correspond to   the younger one. 
Following the same reasoning as for II--A, we favor the younger solution
because high values of $W$(H$\alpha$)  are measured in this region of the galaxy.
This solution yields a very small age: $2.0 \pm 0.8  $ Myr.

The results of the best fits for the other three clusters are given in Table~\ref{tbl05}.
Cluster II--C  yields a single  young solution, while II--D
and II--E have  more complicated outputs; however, their reddenings are 
very similar, as well as their  \rv\ values.

\subsection{Large complexes}

\begin{figure}
\includegraphics[width=0.9\textwidth]{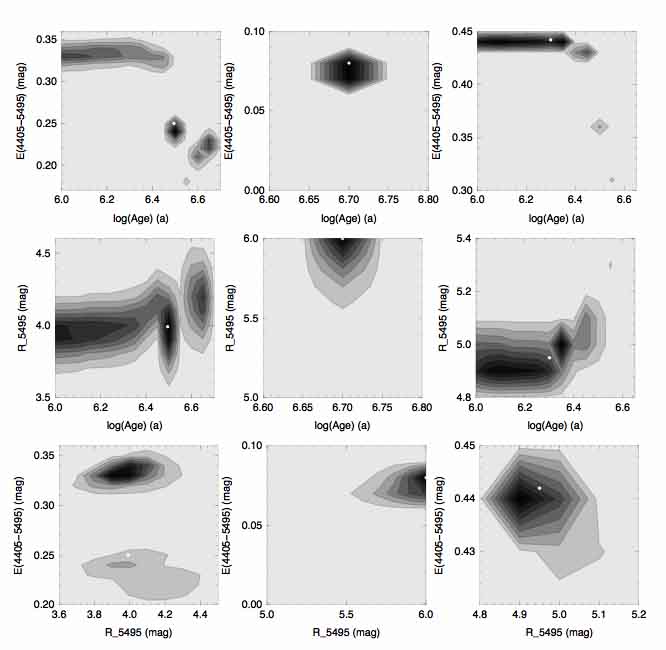}

\caption{ Likelihood contour plots for clusters I--Ds [Left], I--B [Middle], and II [Right]
 provided as output by \cho\ and obtained using \stb\ models.
[Top] Reddening vs. age plot. [Center] Extinction law vs. age. [Bottom] Extinction
law vs. reddening. Age is expressed in years. The white circle marks the most
likely value (mode). }
\label{fig11}
\end{figure}

\begin{figure}
\includegraphics[width=0.9\textwidth]{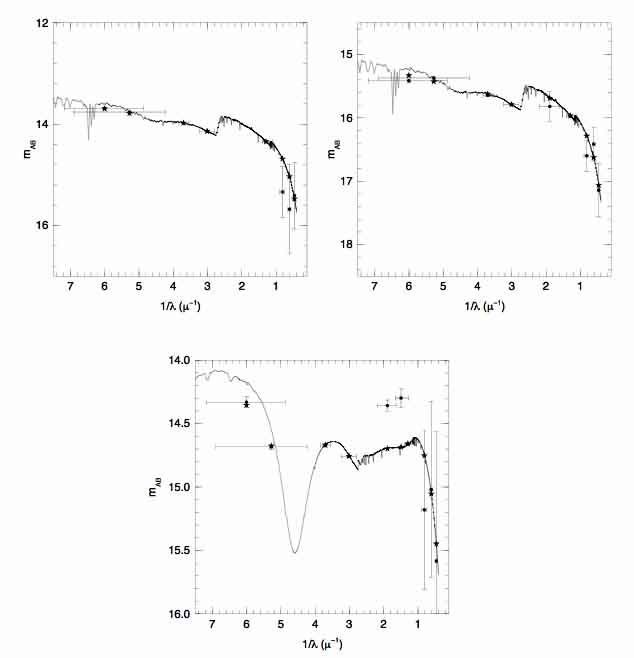}
\caption{SEDs for  the best fits of clusters I--A [Top--left], I--B [Top--right], and II [Bottom], 
based on \stb\ models
of integrated stellar populations. The photometry is shown by the symbols
with error bars  (vertical ones for uncertainties and horizontal ones for 
the approximate length coverage of each filter.) Star symbols indicate
the calculated magnitude of the model SED for each filter.  Magnitudes in 
filters F555W and F702W are displayed in the spectrum of NGC~4214--II, but
they were not considered during the execution of \cho\ because of possible
contamination. See text for details.}
\label{fig12}
\end{figure}

For complexes  I--A, I--B, and II we ran \cho~ 
leaving all three parameters
[ $\log({\rm age}) $, \ecc,  $R_{5495}$ ] unconstrained.
We used the estimated magnitudes from  Table~6 in Paper~I.
Note that for complex~II,
magnitudes F555W and F702W are listed, but they were not considered
for the $\chi^2-$fit. 
The inclusion of the 2MASS magnitudes did not change the output of the code 
significantly in any case.
The best--fit  SEDs are presented in Figure~\ref{fig12}.
The fits of complexes I--A, and I--B are excellent. The F170W magnitude 
is the one that contributes the most to $\chi^2 = 4.25$ in the case of cluster I--B.
Figure~\ref{fig11} shows the contour plots provided as output by \cho\ for cluster I--B.
It is clear from the  plots that for cluster  I--B there is a single solution 
compatible with our photometry. We obtain the same age (5.0 Myr) for both of them 
and, as expected, a low value for the extinction $ E(4405-5495) =  0.07$ mag.
This value is very close to the one obtained for cluster I--As, as we
had expected.
It is important to note that the aperture that we used in both cases is large enough
to include  a part of the galaxy that contains a mixed population of stars. 
Figure~2 in Paper~I  clearly shows both late-- and early--type stars are present
within the apertures, something unexpected for an age of 5.0 Myr.  
Our extinction analysis showed that these large complexes contain stars with  variable
extinction as well. 
This leads to conclude that
the SEDs for clusters I--A and I--B are composite spectra of different complex 
stellar populations. 
Again, the degeneracy in the extinction law for low values of the reddening gives 
high values or \rv.
An estimate of the total mass of these clusters gives:  $(156 \pm 19) \times 10^3$  
$M_{\odot}$ for cluster I--A,
and  $(34 \pm 4) \times 10^3$  $M_{\odot}$ for cluster I--B. All the results
are presented in Table~\ref{tbl06}.

For cluster II we used the largest aperture radius: 200 PC pixels. This means that we 
are considering a mixed population of stars embedded in a gas--rich environment.
As a result, we obtained
a composite spectrum characterized by an age of  $1.9 \pm 0.9$ Myr. Its estimated mass
is    $(9 \pm 3) \times 10^5$  $M_{\odot}$.  The right column in Figure~\ref{fig11}
shows the contour plots obtained for cluster II. The top plot indicates that this cluster is 
younger than cluster I--A, and that it is actually extinct. The middle and 
bottom diagrams imply that the law in the \rv--dependent family of \citet{Cardetal89}
 that best fits  our photometry has a very high value or \rv.

Complex II is expected to house objects with high extinction, since these are located within or very near 
filled \ha\ regions that show little evidence of wind--blown 
or supernova created bubbles \citep{Maizetal98,MacKetal00}.
The extinction map represents very well this part of NGC~4214, showing the
highest values of extinction in the southern part of cluster II--A and between clusters II--B and II--C.
The measured values for the stellar \ecc\ are in good agreement with those obtained from \ha/\hb\ 
\citep{Maizetal98}.
We also calculated the mean value of the stellar extinction in each of 
the clusters  (II--ABCDE) 
 inside complex NGC~4214--II and found excellent agreement with the values obtained 
by fitting \stb\ models to the spatially--integrated magnitudes.
One remarkable  result from   Table~\ref{tbl06} is that  our models yield high values
of \rv\ when we analyze large complexes. When using a large aperture 
one includes a mixed population of stars; we analyzed the \rv\ output for 
the brightest stars in this region and found that individual stars have indeed large values of 
extinction, specially those in cluster II--B. 
It is well known that the value of \rv\ depends upon the environment
along the line of sight. 
A direction  through
 low--density ISM usually has rather low value of extinction (about 3.1). Lines of 
 sight penetrating into  dense molecular clouds like
 Orion, Ophiuchus
 or Taurus  yield  $4 < R_V< 6$  \citep{Math90}.
For example, star Her~36 in M~8 has $R_{5495} = 5.39 \pm 0.09$ \citep{Ariaetal05}.

\section{Summary and conclusions}

\subsection{Extinction}

Using the estimated values of stellar extinction from the output 
of {\sc\footnotesize  CHORIZOS}, we built  an  
extinction map in the field of view where we performed our study.
The most noticeable  characteristic  in this map are the
low values of the extinction scattered throughout the map, with
the exception of  some well defined
regions with high values.   
We compared our extinction map with results provided by 
\cite{Maizetal98} and \citet{Maiz00} who used the Balmer 
ratio (H$\alpha$/H$\beta$) as a tracer of the
reddening.
They  found  
that the reddening in NGC 4214--II is, on average, higher than in NGC 4214--I. 
Our results, derived
from the stellar colors, are in agreement.  
The two main cavities in NGC 4214--I show low 
extinction surrounded by higher values while for NGC 4214--II the extinction is overall higher.
Our extinction 
map traces fairly well  the molecular clouds studied by  
\cite{Waltetal01} which are directly associated to the star--forming regions
in NGC~4214.

Studying the stellar content of seven blue compact galaxies,  
\cite{Faneetal88} found a discrepancy between the  extinction derived from 
the  UV continuum of starburst galaxies and that derived from the Balmer lines.
The extinction derived from UV continuum was systematically lower. 
\cite{Calzetal94}   analyzed $IUE$ UV and optical spectra of 39 starburst and blue 
compact galaxies
in order to study the average properties of dust extinction. They derived 
the UV and optical extinction law
under the hypothesis  that the dust is a screen in front of the source.
The characteristics of the extinction law are different from the 
MW and LMC laws: the overall slope is more gray than the 
MW or LMC slopes, and, most remarkably, the 2\,175 \AA\ dust feature is 
absent within the observational uncertainties.
The different slope explained the differences observed between stellar and nebular extinction.
However, \cite{MasHKunt99} showed that the observed UV to optical SEDs of 
17 starburst  galaxies can be very well reproduced by reddening the 
corresponding synthetic spectra with one of the three extinction laws 
(Galactic, LMC, and SMC) with no need to invoke an additional universal law.  
As proposed by \cite{MasHKunt99} and \cite{MacKetal00}, 
the   effect was merely geometrical: while the continuum 
flux comes from the stars, the nebular emission originates in 
extended regions adjacent to  the original molecular cloud. 
Stellar winds and supernova explosions might wipe out the dust 
from the neighborhood of the massive clusters and concentrate it in 
filaments and dust patches located within the ionizing region.
Depending on the specific geometrical distribution of dust and stars, 
the extinction could affect mainly the nebular gas emission but 
only weakly the stellar continuum. 
This implies that the attenuation law by \cite{Calzetal94}  applies only to regions 
like I--A  in NGC~4214, where the stars have evolved long enough to disrupt the ISM; 
meanwhile, this law should not be applied to NGC~4214--II, because 
the distribution of stars and  dust are co--spatial in this part of the galaxy.

\cite{MacKetal00} analyzed the differential extinction of the gas in NGC~4214
and found that  once 
a cluster becomes older than $\approx 2$ Myr, 
 the stellar components are mostly concentrated in those 
regions where the gas shows very low extinction.
In our current work, we studied the differential extinction of the continuum from the 
stars in the galaxy.
With this research we arrive to the conclusion that  the extinction derived from the stellar continuum  
is similar to the extinction derived via the analysis of 
nebular lines across the galaxy, and we find this to be  true on a pixel by pixel  basis.
The coincidence is fairly good throughout the galaxy.
This confirms the idea advanced by \cite{MasHKunt99} and \cite{MacKetal00} 
that the differences between the   \cite{Calzetal94}  attenuation law and other extinction laws
are caused by  the differences in the spatial distribution
of the ionized gas and the young stellar population.

\subsection{The ratio of blue--to--red supergiants}

We approached the problem of determining the ratio    of blue to
red supergiants (B/R) in NGC~4214. 
We compared our observational results with three sets of theoretical models
calculated with the MC metallicities and arrived at the conclusion that 
the best fit to  our data is the non--rotating,  $Z=0.004  $, Geneva set
  \citep{MaedMeyn01}.
These models predict a B/R value of 24 in the $ 15-20 \,M_{\sun}$ mass--range, and 
we measure  $34 \pm 10$. In the mass--range $ 20-25 \,M_{\sun}$ the
theoretical prediction is 47 and we obtain  $46 \pm 23$.
In both cases, our results agree with  the theoretical ones  within Poisson errors. 
We discussed two caveats in the determination of B/R: the
 stochastic effects due 
to small--number statistics of  RSGs in our sample in NGC~4214, and
the conversion from observed colors to effective temperatures and
bolometric magnitudes. We observe a discrepancy of $350-400$ K in our data
which may be accounted for by 
fitting the  observed optical colors using \cho\ or a similar code 
with  MARCS atmospheres instead of   Kurucz atmospheres.

\subsection{The initial mass function} 
We studied the initial mass function of NGC~4214 following the 
method provided by  \cite{Lequ79} with several  improvements, and 
taking into account four sources of systematic effects
(incompleteness of the data, optimum bin size, mass diffusion, and
unresolved multiple systems)  presented by 
 \cite{Maizetal05}. 
We obtained  a mean value of $ \gamma= -2.83$  for the IMF slope
of NGC~4214 in a continuous star--formation scenario.
Since the correction for unresolved objects is uncertain  (given the large distance to
NGC~4214), and knowing that unresolved objects produce a systematic 
flattening of the IMF slope, we conclude that 
the real value of the slope would be steeper than $ \gamma= -2.83$. 
We expect the ``real'' IMF slope to be shallower than the one
provided by  \cite{Chanetal05} ($ \gamma = -3.5$) and 
steeper than our calculated value.
Our estimation is   closer to 
the one calculated by 
\cite{MasHKunt99} ($ \gamma = -3.0$) for the sum of field and clusters
in NGC~4214.
Some OB associations in the Magellanic Clouds and the Milky Way
\citep{Masse98} 
have comparable values of the IMF slope 
to the one  that we obtained for NGC~4214.

\subsection{Clusters}
 
We searched for the best fits of \stb\ models to photometric colors
of 13 clusters in NGC~4214.
We chose four unresolved compact clusters, three large complexes and six
small resolved clusters.

We find that the best models that are compatible with
 our data of cluster I--As are those with a MC--like extinction law. 
These models predict that  cluster I--As is a young  ($4.0 \pm 0.6$ Myr) 
massive  ($\approx 27\,000 M_{\odot}$) star cluster
with low ( $E(4405-5495) = 0.04 $ mag ) extinction.
Our study is consistent with  the presence of WR stars in this cluster, with 
a number close to  the value previously obtained by \cite{Leitetal96}
and \cite{Maizetal98}.
\cite{MacKetal00}  estimated the equivalent widths of the Balmer lines
$W$(H$\alpha$) 
 using three extinction corrections for all
the clusters in NGC~4214, in order to restrict the clusters ages.
For clusters NGC~4214--I--A and  I--B they estimate an age in the range between 
4 and 5  Myr, which
agrees with our photometric method. 
Their  age estimate for cluster  I--Ds is 7 Myr, while we obtain a much smaller
value: $2.6 \pm 1.5$ Myr.
Cluster I--Es provides two very different solutions, but with 
 the current data
we prefer not to favor any of the solutions over the other. 

\cite{MacKetal00}   find equivalent widths of \ha\
very close or higher than 1000 \AA\ for all the knots in NGC~4214--II,
except for II--C, indicating that this large complex  is very young. We consistently obtain
very small ages for all the clusters within  NGC~4214--II, including cluster II--C. 
The overall value of the $W$(H$\alpha$) for NGC~4214--II is quite lower than 1000 \AA.
\cite{MacKetal00}  explain that this may be due to the existence of an underlying population.
They took this effect into account
and estimated the  value $2.5-3.0$ Myr for the age of   cluster  II. Our 
age estimate is $1.9 \pm 0.9$ Myr which agrees with their value.
The  $W$(H$\alpha$)  for NGC~4214--I is in general lower than those
of  for NGC~4214--II, and these authors suggest an average age in the
range $3.0-4.0 $ Myr for cluster I.

Our analysis for clusters IIIs and IVs includes six photometric colors
covering wavelengths from the UV to the IR. This allowed us to find restrictions 
to the age and mass of these clusters.
Using several arguments (extinction values, number of red supergiants, image appearance)
we suggest that the older solutions should be preferred 
over the younger ones in the case of these two clusters.
\cite{Billeetal02}   study clusters IIIs and IVs as part of a survey of compact star
clusters in nearby galaxies which include NGC~4214. Unfortunately, their aperture photometry 
includes  errors such as not taking into consideration the contamination effect 
for filter F336W and the conversion of HST flight magnitudes
to the Johnson--Cousins system, which  is not recommended 
because of  their limited precision \citep{Gonzetal03,deGretal05}. 
Using an evolutionary track taken from the  \stb\ models on a color--color diagram,
they infer age estimates to these clusters. The evolutionary track on this diagram
is clearly degenerate, and shows 
several possible solutions for the age of these clusters.
They used only two photometric colors, and this  translates
into a very poor determination of the clusters age  \citep{deGretal03a}. 
Using  the photometry produced by   \cite{Billeetal02} and the cluster 
evolutionary models of  \cite{BruzChar03},    \cite{Larsetal04} infer the age 
and mass of clusters IIIs and IVs. The mass is inferred  using a dynamical method
for clusters in virial equilibrium.

We find that  NGC~4214--IIIs and --IVs  are compact old massive clusters with age $> 100$ Myr.
Ten other clusters (one unresolved compact cluster,  three large complexes and six
small resolved clusters) yield ages  $< 10$ Myr.
Cluster luminosity functions and color distributions are the most important tools in
the study of cluster populations in nearby galaxies.
The use of individual cluster spectroscopy, acquired with 8--m--class telescopes
is very time consuming, because observations of large numbers of clusters
are needed in order to obtain statistically significant results.
Multipassband  imaging is a useful alternative. \cite{deGretal03a} analyze 
the systematic uncertainties in age, extinction, and metallicity determinations
for young stellar clusters, inherent to the use of broad--band, integrated colors. 
They studied clusters within NGC~3310 and  found that  red--dominated 
passband combinations result in significantly different
age solutions, while blue--selected passband combinations tend to result in age 
estimates that are slightly skewed towards lower ages. 
Their advise is to use at least four filters including both blue and red optical
passbands. This choice leads to the most representative age distribution. 
See also \cite{Andeetal04} and  \cite{deGretal05}.

For our  cluster analysis in NGC~4214, we employed at least five  filters in each case,
and we determined all the free parameters individually for
each cluster as suggested by \cite{deGretal03b}.
The accuracy to which the ages can be estimated depends on the number of 
different broad--band filters and, crucially, on the actual wavelengths range covered by 
the observations. 
We found some degeneracies that could have been disentangled, 
provided that we had a measurement of a
magnitude near the right side of  the 
Balmer discontinuity  such as F439W (WFPC2 $B$).
However, we are confident with the age estimate of our sample of clusters
in NGC~4214.

Employing multicolor images of the Antennae galaxies, \cite{Falletal05}
studied the age distribution of the  population of star clusters.
They estimated the cluster ages, by fitting SEDs from the \cite{BruzChar03} models and found
the age distribution declines steeply, starting at very young ages. The median age of the clusters
is  $\sim10$ Myr. According to their study, after $\sim10$ Myr, the surface brightness of the clusters
would be 5 mag fainter than initially (at $\sim1$ Myr), and therefore
the cluster would disappear among the statistical fluctuations in the foreground 
and background 
of field stars. They call this effect the ``infant mortality" of clusters.
Our sample of clusters in NGC~4214 includes 10 objects younger than  10 Myr,
with NGC~4214--IA and --IB already showing effects of disruption. 
What could cause the disruption of the clusters?
\cite{Falletal05} claim that the momentum output 
from massive stars comes in the form of ionizing radiation, stellar winds, jets, and 
supernovae, and all these processes could easily remove much of the ISM from the
protocluster leaving the stars within it gravitationally unbound. 
Another explanation is provided by \cite{Claretal05} who performed
simulations and showed that unbound giant
molecular clouds manage to form a series of star clusters and disperse 
in $\sim10$ Myr, making them transient features. At later
ages they would not be recognized as clusters.
This result is further confirmed  by several empirical and theoretical 
studies. See   \cite{LadaLada03} for a review and the references therein.
\cite{Lameetal05a}  present a simple analytical description
of the disruption star clusters in a tidal field, and found that about half of the
clusters in the solar neighborhood become unbound within about 
10 Myr.

\subsection{Implications}
Nearby galaxies provide ideal laboratories  to test how stars form, how 
star formation is triggered, and details of how galaxies assembled. 
The study of nearby galaxies like NGC~4214 is rather important because 
most of the information that we infer about  high--redshift  galaxies relies on
what we observe in galaxies in the local universe. Therefore, understanding
nearby galaxies is of extreme importance to comprehend what
is going on in more distant ones.
NGC~4214 is a low--metallicity galaxy, and this gave us the possibility
to study the physical conditions   in an environment 
of current astrophysical interest.
In particular, we could infer that the B/R supergiant ratio
is close to the one of the SMC
and that its IMF is steeper that Salpeter. 


\acknowledgments
We want to thank the referee, Dr. Richard de Grijs, whose constructive and detailed 
report helped immensely to  improve the quality of this paper.
We also want to thank Rupali Chandar and Claus Leitherer  for useful comments on the
original draft of this work.

Support for this work was provided by NASA through grants GO--06569.01--A, 
GO--09096.01--A, GO--09419.01--A, and 
AR--09553.02--A from the Space Telescope Science Institute, Inc., under NASA contract 
NAS5--26555.
This research has made use of the VizieR catalogue access tool, CDS, Strasbourg, France.
This publication makes use of data products from the Two Micron All Sky Survey, which
 is a joint project of the University of Massachusetts and the Infrared Processing and 
 Analysis Center/California Institute of Technology, funded by the National Aeronautics 
 and Space Administration and the National Science Foundation.

Facilities: \facility{HST(WFPC2)}, \facility{HST(STIS)}.


\clearpage

\clearpage

\clearpage
 
\clearpage

    
 \clearpage

\clearpage

     \clearpage

\clearpage
 

\clearpage


\clearpage


\clearpage

\clearpage

\clearpage

\begin{table}[htbp]
\caption{Comparison between theoretical  and observational values of 
the ration B/R of the number of blue supergiants and red supergiants.
 \label{tbl01}
\vspace{5mm}}
\begin{tabular}{@{} c|c|c|c|c|c|c @{}}   \tableline\tableline
           &\multicolumn{2}{c|}{SMC  $v_{ini} \tablenotemark{a}=0$}      &\multicolumn{2}{c|}{SMC  $v_{ini} =300$}    
           & \multicolumn{2}{c}{LMC  $v_{ini} =0$}    {\rule [0mm]{0mm}{5mm} }    \\   
  \raisebox{1.8ex}[0pt]{mass range}         & theory  &   obs   &   theory  &   obs   & theory  &   obs  {\rule [-3mm]{0mm}{0mm} }       \\   \tableline
   $   15-20 $   & 24 & $34 \pm 10$ & 0.4 & $29 \pm 10$& 1.0 & $13 \pm 4$  {\rule [0mm]{0mm}{5mm} }\\
    $  20-25 $   & 47 & $46 \pm 23$ & 0.4 & $45 \pm 31$& 2.7 & $67  \pm 66${\rule [-3mm]{0mm}{0mm} }\\ \tableline
\end{tabular}
\tablenotetext{a}{  $v_{ini}$ is the initial rotational velocity of the theoretical model, in km s$^{-1}$.  }

\end{table}

\clearpage

\begin{deluxetable}{lccccccccccc} 
\tabletypesize{\scriptsize}
\tablecaption{ IMF slope values  with their errors for the three lists calculated for several  values
of the low--mass end  ($M_{low}$) and $M_{up} = 100 M_{\odot}$. These results would apply
to a continuous star--formation region. \label{tbl02}}
\tablewidth{0pt}
\tablehead{  \colhead{} 
& \multicolumn{3}{c}{ I--A +  I--B +  II}  &  \colhead{}
& \multicolumn{3}{c}{ Other regions} &  \colhead{}
& \multicolumn{3}{c}{NGC~4214}  \\
\cline{2-4}\cline{6-8}\cline{10-12} \\    
\colhead{ $M_{low} [M_{\odot}]$}    &  \colhead{$\gamma_{fit}$} &  \colhead{$\gamma_{real}$}   & \colhead{$N$}  &\colhead{}  &
\colhead{$\gamma_{fit}$} &  \colhead{$\gamma_{real}$}  &  \colhead{$N$}  &\colhead{} &
 \colhead{$\gamma_{fit}$} &  \colhead{$\gamma_{real}$}  &   \colhead{$N$}    }
\startdata
  15 &    -2.17 &   $ -2.67 \pm     0.08 $&  752 &  &  -2.15 &  $  -2.64 \pm     0.06 $& 1680 &  &  -2.15 & $   -2.64 \pm      0.05  $ & 2432\\
   16 &    -2.20 &  $  -2.61 \pm     0.09 $&  655 &   & -2.28 &  $  -2.72 \pm     0.06 $& 1494 &  &  -2.26 & $   -2.70 \pm      0.05 $  & 2149\\
   17 &    -2.24 &    $-2.60 \pm     0.10 $&  573 &   & -2.43 &  $  -2.83 \pm    0.07 $& 1357 &  &  -2.39 & $   -2.78 \pm         0.06 $  & 1930\\
   18 &    -2.21 &   $ -2.51 \pm     0.10 $&  501 &   & -2.59 &$    -2.95 \pm     0.08 $& 1216 &  &  -2.48 & $   -2.82  \pm        0.06  $ & 1717\\
   19 &    -2.30 &   $ -2.57 \pm     0.11 $&  449 &  &  -2.72 & $   -3.03 \pm     0.08$ & 1098 &  &  -2.56 &$    -2.86 \pm     0.06  $ & 1547 \\
   20 &    -2.36 &   $ -2.60 \pm     0.12$ &  397 &  &  -2.71 &$    -2.98 \pm     0.09 $&  961 &  &  -2.58 & $   -2.83 \pm        0.07$   & 1358   \\
\enddata
\end{deluxetable}


\begin{deluxetable}{llccc} 
\tabletypesize{\scriptsize}
\tablecaption{IMF slope values  with their errors for  regions   I--A,  I--B, and  II. 
These results would apply
to a burst star--formation region.   \label{tbl03}}
\tablewidth{0pt}
\tablehead{  \colhead{} &\colhead{} 
& \multicolumn{3}{c}{ I--A +  I--B +  II}    \\
\cline{3-5}\\  
\colhead{ $M_{low} [M_{\odot}]$}    & \colhead{ $M_{up} [M_{\odot}]$}    & \colhead{$\gamma_{fit}$} &  \colhead{$\gamma_{real}$}    & \colhead{$N$}  }
\startdata
   15 & 100 &   -3.30 &$    -4.41 \pm     0.09 $&  752   \\
   16 &  100 &  -3.25 &  $  -4.06 \pm     0.10 $&  655  \\
   17 & 100 &   -3.24 & $   -3.87 \pm    0.10 $&  573  \\
   18 &  100 &  -3.14 & $   -3.61 \pm    0.11 $&  501 \\
   19 & 100 &   -3.20 & $   -3.59 \pm     0.12 $&  449  \\
   20 & 100 &   -3.23 & $   -3.56 \pm     0.13 $&  397   \\
      40 & 100 &   -2.89 & $   -2.97 \pm     0.37 $&  85   \\
   20 & 40 &   -3.73 & $   -4.41 \pm     0.28 $&  312   \\

\enddata
\end{deluxetable}

\clearpage
\begin{deluxetable}{llllllllllll}
\tabletypesize{\scriptsize}
\rotate
\tablecolumns{12}
\tablewidth{0pt}
\tablecaption{Results from  \cho ~fitting for clusters I--As, I--Es, IIIs, and IVs. The mass was estimated using a  Kroupa IMF \citep{Krou02}.
 The number of stars is obtained from \stb\ models.
\label{tbl04} }
\tablehead{
 \multicolumn{1}{l}{Parameter}     
&  \multicolumn{2}{c}{Cluster I--As} &  \colhead{}  
&  \multicolumn{2}{c}{Cluster I--Es} &  \colhead{}
&  \multicolumn{2}{c}{Cluster IIIs}  &  \colhead{}   
&   \multicolumn{2}{c}{ Cluster IVs}     \\[-00mm]
\cline{2-3}\cline{5-6}\cline{8-9}\cline{11-12} \\[-2mm]
 \colhead{}
& \multicolumn{1}{c}{LMC2}    
&  \multicolumn{1}{c}{SMC}    
& \colhead{} 
&  \multicolumn{1}{c}{Young}    
& \multicolumn{1}{c}{Old}    
& \colhead{}   
& \multicolumn{1}{c}{Young}    
&  \multicolumn{1}{c}{Old}    
& \colhead{} 
&  \multicolumn{1}{c}{Young}       
& \multicolumn{1}{c}{Old}     
}
\startdata
Age (Myr)  &      $4.0 \pm 0.6$ &  $4.2 \pm 0.6$ &  &   $ 7.1\pm0.1$    & $189\pm53$   
           &        &      $15.8\pm 0.2$ &  $168 \pm 61$ & &   $15.8 \pm 0.0$    & $150 \pm 34$    \\ 
\ecc ~(mag)  &   $0.05 \pm 0.01$    &    $0.04 \pm 0.01$  &   &  $0.46\pm0.01$   & $0.32\pm0.03$   
            &          &   $0.44 \pm 0.01$    &    $0.26 \pm 0.03$  &   &  $0.22\pm0.01$   & $0.06\pm0.03$   \\ 
Extinction law  / \rv  &   LMC2    &  SMC    & &     $4.21\pm0.25$  & $5.72\pm0.26$   
             &            &   $3.95\pm0.20$     &  $4.98\pm0.60$   &   &     $3.39\pm0.52$  & $4.35\pm1.12$   \\ 
Mass ($10^3 M_{\odot} $)   & $ 27\pm   4  $ & $ 27 \pm  4  $ & &    $6\pm1$     &  $129\pm26$   
                                &             & $ 134\pm   18  $ & $ 626 \pm   175  $ & &    $27\pm4$     &  $114\pm20 $   \\ 
$\chi^2  $ per degree of freedom   &    1.33   &   0.91  &  &  2.58 &  2.19 & &    0.39   &   0.66   & &  1.77  &  2.18   \\ 
O+B stars, types I and II  &     11  &  11 &   &2 &  0  &   &     31  &  0 &   &7 &  0  \\ 
K+M stars, types I and II  &   0    & 0  &  &  1 & 9   &   &   6   &  45  & &  1 & 26 \\ 
WR stars  & 1     & 1  & &   0 &  0   &   & 0     & 0&  &  0 &  0  \\[0mm]  
\enddata
\end{deluxetable}

\clearpage

\begin{deluxetable}{lllllll}
\tabletypesize{\scriptsize}
\tablecolumns{7}
\tablewidth{0pt}
\tablecaption{Results from  \cho ~fitting for resolved clusters.The mass was estimated using a  Kroupa IMF \citep{Krou02}.
\label{tbl05} }
\tablehead{
\multicolumn{1}{l}{Parameter}   & \multicolumn{1}{c}{Cluster I--Ds}    
&  \multicolumn{1}{c}{Cluster II--A}    
&  \multicolumn{1}{c}{Cluster II--B}    
&  \multicolumn{1}{c}{Cluster II--C}    
& \multicolumn{1}{c}{Cluster II--D}    
& \multicolumn{1}{c}{Cluster II--E}    
}
\startdata
Age (Myr)  &      $2.6 \pm 1.5$ &  $3.1 \pm 1.4$   &   $ 2.0\pm0.8$    & $1.7\pm0.6$   &      $4.0\pm 4.0$ &  $3.1 \pm 1.4$   \\ 
\ecc ~(mag)  &   $0.28 \pm 0.05$    &    $0.35 \pm 0.08$     &  $0.35\pm0.01$   & $0.28\pm0.01$ & $0.22 \pm 0.16$    &    $0.20 \pm 0.05$     \\  
Extinction law / \rv   &       $4.05\pm0.19$  & $3.79\pm0.33$ &   $4.11\pm0.11$     &  $4.82\pm0.13$      &     $4.05\pm0.91$  & $2.92\pm0.43$   \\ 
Mass ($10^3 M_{\odot} $)   & $ 20 \pm   9  $ & $ 34 \pm   23  $  &    $43\pm10$     &  $63\pm14$ & $ 7\pm   5  $ & $ 7 \pm 3  $     \\ 
$\chi^2  $ per degree of freedom   &    1.31   &   0.41  &     2.02 &  2.53 &    0.15   &   0.22   \\ 
\enddata
\end{deluxetable}


\begin{deluxetable}{llll}
\tabletypesize{\scriptsize}
\tablecolumns{4}
\tablewidth{0pt}
\tablecaption{Results from  \cho ~fitting for large complexes.The mass was estimated using a  Kroupa IMF \citep{Krou02}.
\label{tbl06} }
\tablehead{
\multicolumn{1}{l}{Parameter}   &
\multicolumn{1}{c}{Cluster I--A}    
&  \multicolumn{1}{c}{Cluster I--B}    
&  \multicolumn{1}{c}{Cluster II}    
}
\startdata
Age (Myr)  &      $5.0 \pm 0.0$ &  $5.0 \pm 0.0$   &   $ 1.9\pm0.9$       \\ 
\ecc ~(mag)  &   $0.07 \pm 0.01$    &    $0.07 \pm 0.01$     &  $0.43\pm0.03$  \\
Extinction law / \rv   &       $5.88\pm0.16$  & $5.81\pm0.19$ &   $4.99\pm0.11$    \\  
Mass ($10^3 M_{\odot} $)   & $ 156 \pm   19  $ & $ 34 \pm   4  $  &    $923\pm331$    \\
$\chi^2  $ per degree of freedom   &    0.95   &   4.25  &   0.21     \\ 
\enddata
\end{deluxetable}

\end{document}